\documentclass[unsortedaddress,preprint,aps,showpacs]{revtex4}
\usepackage{graphicx}
\usepackage{dcolumn}
\usepackage{amsmath}
\usepackage{amssymb}
\usepackage{amsfonts}
\usepackage{bm}
\usepackage{color}
\newcommand{\be}{\begin{equation}}
\newcommand{\ee}{\end{equation}}
\newcommand{\ba}{\begin{eqnarray}}
\newcommand{\ea}{\end{eqnarray}}
\begin{document}
\title{A fingerprint of surface-tension anisotropy \\
in the free-energy cost of nucleation}
\author{Santi Prestipino$^1$\footnote{Corresponding author. E-mail:
sprestipino@unime.it}, Alessandro Laio$^2$\footnote{E-mail: laio@sissa.it},
and Erio Tosatti$^{2,3}$\footnote{E-mail: tosatti@sissa.it}}
\affiliation{
$^1$ Universit\`a degli Studi di Messina, Dipartimento di Fisica e di
Scienze della Terra, Contrada Papardo, I-98166 Messina, Italy \\
$^2$ International School for Advanced Studies (SISSA) and UOS Democritos,
CNR-IOM, Via Bonomea 265, I-34136 Trieste, Italy \\
$^3$ The Abdus Salam International Centre for Theoretical Physics (ICTP),
P.O. Box 586, I-34151 Trieste, Italy}
\date{\today}
\begin{abstract}
We focus on the Gibbs free energy $\Delta G$ for nucleating a droplet
of the stable phase (e.g. solid) inside the metastable parent phase
(e.g. liquid), close to the first-order transition temperature. This
quantity is central to the theory of homogeneous nucleation,
since it superintends
the nucleation rate. We recently introduced a field theory describing
the dependence of $\Delta G$ on the droplet volume $V$, taking into
account besides the microscopic fuzziness of the droplet-parent interface,
also small fluctuations around the spherical shape whose effect, assuming
isotropy, was found to be a characteristic logarithmic term.
Here we extend this theory, introducing the effect of anisotropy
in the surface tension, and show that in the limit of strong anisotropy
$\Delta G\left(V\right)$ once more develops a term logarithmic on $V$,
now with a prefactor of opposite sign with respect
to the isotropic case. Based on this result, we argue that the geometrical
shape that large solid nuclei mostly prefer could be inferred from the
prefactor of the logarithmic term in the droplet free energy,
as determined from the optimization of its near-coexistence profile.
\end{abstract}
% 64.60.qe : General theory and computer simulations of nucleation
% 68.03.Cd : Surface tension and related phenomena
% 68.35.Md : Surface thermodynamics, surface energies
\maketitle

\section{Introduction}
\setcounter{equation}{0}
\renewcommand{\theequation}{1.\arabic{equation}}

When a homogeneous, defect-free bulk system is brought across a first-order
phase boundary, it may survive in its metastable state even for a long time,
until the stable phase spontaneously nucleates~\cite{Kelton,Kashchiev}.
The nucleation process has attracted much attention over the years,
both from a fundamental point of view as well as for its great practical
interest. To mention but one example, a better control of crystal
nucleation in protein solutions could help hinder protein condensation
which is at the heart of several human pathologies~\cite{Vekilov}.
Thermal fluctuations continuously sprout droplets of the stable phase
inside the metastable mother phase. Small droplets dissolve, for the
gain in volume free energy fails to compensate the loss in surface
free energy. Occasionally, a droplet is sufficiently large that it
is favorable for it to grow. Once this happens, the solid nucleus
expands until the whole liquid crystallizes. Quenching the system
deeper and deeper lowers the nucleation barrier until the point where
the barrier vanishes (kinetic spinodal limit). Beyond this threshold,
nucleation ceases and the phase transition occurs through spinodal
decomposition and coarsening (i.e., uniformly throughout the material).
Classical nucleation theory (CNT)~\cite{Volmer,Farkas,Becker} provides
the simplest theoretical framework in which the initial stage of
the phase transformation can be described. In this theory, an isolated
droplet is schematized, regardless of its size, as a sphere of bulk
solid, separated from the liquid by a sharp interface with a constant
free-energy cost per unit area $\sigma$ (``capillarity approximation'').
This gives rise to a (Gibbs) free-energy difference between the supercooled
liquid system with and without a solid cluster, that is
\be
\Delta G(V)=-\rho_{s}|\Delta\mu|V+(36\pi)^{1/3}\sigma V^{2/3}\,,
\label{1-1}
\ee
where $V$ is the cluster volume, $\Delta\mu<0$ is the difference
in chemical potential between solid and liquid, and $\rho_{s}$ is the
bulk-solid number density. The droplet grows if it exceeds a critical size
$V^{*}$ corresponding to the maximum
$\Delta G$ ($\equiv\Delta G^{*}$), which thus provides the activation
barrier to nucleation~\cite{Bagdassarian}. 

The cluster free energy $\Delta G(V)$ can be accessed numerically via the
statistics of cluster size, through which the validity of Eq.\,(\ref{1-1})
for specific model interactions can be directly tested.
We recently showed~\cite{Prestipino}
that the accuracy of CNT is less than satisfactory in estimating the
size probability distribution of clusters, especially the smaller
ones, implying that interface-tension estimates based on the use of
CNT are systematically in error. We then proposed a more detailed field
theory of the nucleation barrier, based on the assumption that clusters
are soft and not sharp, and can deviate mildly from the spherical
shape (``quasispherical'' approximation).
If the solid-liquid interface tension is taken to be isotropic,
the volume dependence of the Gibbs free energy of a cluster is of the
Dillmann-Meier form~\cite{Dillmann},
\be
\Delta G(V)=-\rho_s|\Delta\mu|V+AV^{2/3}+BV^{1/3}+C-\frac{7}{9}k_{B}T\ln
\frac{V}{a^3}\,,
\label{1-2}
\ee
where $A,B$, and $C$ can all be expressed as explicit functions of the
``microscopic'' parameters entering a Landau free energy, and $a$ is a
microscopic length.
It turned out that the numerical profiles of $\Delta G$ in a few test
cases and at various supersaturations are better reproduced by
this theory.

Here we critically reconsider the most severe assumption made 
in that derivation, namely the isotropy of the solid-liquid interface
tension. We show that the theory introduced in Ref.\,\cite{Prestipino}
can be extended relaxing this important approximation, and that the results
change. Starting once
again from a Landau-like theory, we derive an interface Hamiltonian,
that allows estimating the probability of observing a cluster of any
shape and size. The angular dependence of the interface tension is taken into
account by terms that depend on the local orientation of the cluster
surface. Within this framework, we calculate $\Delta G(V)$ in the limit
of strong surface anisotropy and compare it with the isotropic case.
For large anisotropy, the cluster free energy still retains at large size
a logarithmic term, however with a prefactor of opposite
sign to the isotropic one. On account of this, we suggest that the
nominal shape of large solid nuclei could be guessed from the optimization
of the actual $\Delta G(V)$ close to coexistence.
Looking for a numerical exemplification, we conducted 3D
Monte Carlo simulations of the Ising model extracting $\Delta G(V)$
for clusters of variable size $V$, at various distances from coexistence.
Although we could not really attain sizes where the anisotropic shape
effects are heavy, we do detect evidence that the sign is as expected
for large anisotropy.

The paper is organized as follows. We start in Section II by relaxing
the approximation of an infinitely sharp cluster interface, with the
introduction
of a Landau free energy. From that, an effective sharp-interface
Hamiltonian is derived in Section III, as an intermediate step to
building up a field theory for isotropic surfaces where small
shape fluctuations are allowed (Section IV.A). Eventually, this leads
to a modified-CNT expression of $\Delta G(V)$. In Section IV.B, the
issue of interface anisotropies is addressed, and we show by examples
how the dependence of the interface free energy on the local surface
normal affects the formation energy of a large cluster. Next, in
Section V, we check our theory against old and fresh Monte Carlo
simulation data for the nucleation barrier to magnetization reversal
in the 3D Ising model above the roughening temperature. While confirming
that CNT is not generally adequate to fit
the numerical $\Delta G(V)$ data, this analysis also gives a
quantitative measure of the errors made with CNT and demonstrates their
cancellation in the more general theory. Finally, our conclusions are
presented in Section VI.

\section{Diffuse interface: Landau theory}
\setcounter{equation}{0}
\renewcommand{\theequation}{2.\arabic{equation}}

The main assumption behind CNT is that of a sharp and
spherical cluster surface.
A way to relax this approximation is through the introduction of a
scalar, non-conserved order-parameter (OP) field $\phi({\bf x})$
(``crystallinity'') which varies smoothly from one phase to the other.
Hence, the solid-liquid interface becomes diffuse in space, even though
only on a microscopic scale. In practice, $\phi$ may be thought of as the
local value of the main Fourier coefficients of the crystal-periodic
one-body density $n({\bf x})$, i.e., those relative to the reciprocal-lattice
vectors which are closest in modulus to the point where the liquid structure
factor reaches its maximum~\cite{Shih}.
Otherwise, $\phi$ may be identified with the parameter discriminating
between solid and liquid in an ansatz like
\be
n({\bf x})=\left(\frac{\phi}{\pi}\right)^{3/2}\sum_{\bf R}
e^{-\phi({\bf x}-{\bf R})^2}=\rho_s\sum_{\bf G}e^{-G^2/(4\phi)}
e^{i{\bf G}\cdot{\bf x}}\,,
\label{2-1}
\ee
assuming a specific crystal symmetry and an overall number density $\rho_s$.

Across the solid-liquid interface, $\phi$ is no longer constant and,
for a system with short-range forces, the
thermodynamic cost of the interface may be described through the free-energy
functional~\cite{Cahn,Cahn2,Fisher,Harrowell,Shen}
\be
{\cal G}[\phi;\hat{\bf n}]=
\int{\rm d}^3x\,\left\{\frac{c(\hat{\bf n})}{2}(\nabla\phi)^2+
\frac{\kappa(\hat{\bf n})}{2}(\nabla^2\phi)^2+g(\phi({\bf x}))\right\}\,,
\label{2-2}
\ee
where $c,\kappa>0$ are stiffness parameters dependent on
the interface orientation as defined by the unit normal $\hat{\bf n}$ and
$g(\phi)$ is the specific Landau free energy of the homogeneous system,
taken the bulk liquid as a reference.
In Eq.\,(\ref{2-2}), besides the customary square-gradient
term, also a square-laplacian term appears. This is the next-to-leading
isotropic term in the gradient expansion of the Landau free-energy
density~\cite{Kardar}. Even though being a fourth-order gradient term,
it is however only second-order in the order parameter, and this places
it on the same footing as the square-gradient term (hence, potentially
relevant). We shall see below that, without such a term, the bending
rigidity (i.e., the coefficient of $H^2$ in Eq.\,(\ref{3-15}) below)
would simply be zero. Below the melting temperature $T_m$,
$g$ shows, besides the liquid minimum, also a second and deeper solid minimum.
Exactly at coexistence, the two minima are equal, falling at
$\phi_-=\phi_{s0}$ in the bulk solid and at $\phi_+=0$ in the bulk liquid,
which means that $g(\phi_{s0})=g(0)=0$ while $g(\phi)>0$ otherwise.

When boundary conditions are applied such that $\phi\rightarrow\phi_\pm$
for $z\rightarrow\pm\infty$, a planar interface orthogonal to $z$
is forced to appear in the system. The corresponding OP profile
is the stationary solution $\phi_0(z;\hat{\bf n})$ of (\ref{2-2}) that
satisfies the boundary conditions:
\be
c(\hat{\bf n})\phi_0^{\prime\prime}-
\kappa(\hat{\bf n})\phi_0^{\prime\prime\prime\prime}=
\frac{{\rm d}g}{{\rm d}\phi}(\phi_0;T=T_m)\,,\,\,\,\,\,\,{\rm with}\,\,\,
\phi_0(-\infty)=\phi_{s0}\,\,\,{\rm and}\,\,\,\phi_0(+\infty)=0\,.
\label{2-3}
\ee
From now on, we simplify the notation by dropping any reference to
$\hat{\bf n}$ in $c,\kappa$, and $\phi_0$.
Equation (\ref{2-3}) can be simplified by multiplying both sides by
$\phi_0'(z)$ and integrating by parts. We thus arrive at a new boundary
value problem:
\be
\kappa\phi_0'\phi_0'''=\frac{c}{2}\phi_0'^2+\frac{\kappa}{2}\phi_0''^2-
g(\phi_0)\,,\,\,\,\,\,\,{\rm with}\,\,\,
\phi_0(-\infty)=\phi_{s0}\,\,\,{\rm and}\,\,\,\phi_0(+\infty)=0\,.
\label{2-4}
\ee
Obviously, ${\cal G}[\phi_0]$ represents the free-energy cost of the interface
at $T=T_m$.

At temperature below coexistence, the absolute minimum of $g(\phi)$ falls at
$\phi=\phi_s>0$ for $\Delta T\equiv T-T_m<0$. This can be described by
\be
g(\phi)=c_2\phi^2+c_3\phi^3+c_4\phi^4+\ldots
\label{2-5}
\ee
with $c_2=c_{20}+c'_{20}\Delta T$ ($c_{20},c'_{20}>0$), all other
$c_n$ coefficients being constant.

For the remaining part of this Section, we will assume that $c$ and $\kappa$
do not depend on $\hat{\bf n}$. Under this condition, a large solid cluster
can be assumed to be spherical, with a OP profile described by
$\phi_0(r-R)$~\cite{Fisher}, provided the center of $\phi_0(z)$ is at $z=0$.
From this ansatz, in \cite{Prestipino} we derived 
an expression for the cluster free energy,
\be
\Delta G(R)=4\pi R^2\sigma^{\rm L}\left(1-\frac{2\delta^{\rm L}}{R}+
\frac{\epsilon^{\rm L}}{R^2}\right)-\frac{4}{3}\pi R^3\rho_s|\Delta\mu|\,,
\label{2-6}
\ee
in terms of quantities ($\sigma^{\rm L},\delta^{\rm L},\epsilon^{\rm L}$)
which depend linearly on the supersaturation $|\Delta\mu|\propto|\Delta T|$.
Equation (\ref{2-6}) resembles the CNT expression, Eq.\,(\ref{1-1}), with
the crucial difference that the interface free energy is now a
function of both $R$ and $T$:
\be
\sigma(R;T)=\sigma^{\rm L}\left(1-\frac{2\delta^{\rm L}}{R}+
\frac{\epsilon^{\rm L}}{R^2}\right)\,.
\label{2-7}
\ee
Exactly of this form is the tension of the equilibrium
interface between a liquid droplet and the vapour background in the
Lennard-Jones model, as being extracted from the particle-number histogram
in grand-canonical simulations of samples of increasing size~\cite{Block}.
At coexistence, the solid-liquid interface tension and the Tolman
length~\cite{Tolman} are given by:
\ba
&& \sigma_m\equiv\sigma^{\rm L}(T_m)=
\int_{-\infty}^{+\infty}{\rm d}z
\left[c\phi_0^{\prime\,2}(z)+2\kappa\phi_0^{\prime\prime\,2}(z)\right]\,;
\nonumber \\
&& \delta_m\equiv\delta^{\rm L}(T_m)=
-\frac{\int_{-\infty}^{+\infty}{\rm d}z\,z
\left[c\phi_0^{\prime\,2}(z)+2\kappa\phi_0^{\prime\prime\,2}(z)\right]}
{\int_{-\infty}^{+\infty}{\rm d}z
\left[c\phi_0^{\prime\,2}(z)+2\kappa\phi_0^{\prime\prime\,2}(z)\right]}
\label{2-8}
\ea
A nonzero $\delta_m$ occurs if and when $\phi_0(z)$ is
{\it asymmetric} around zero, as is generally the case for the interface
between phases of a different nature (see Appendix A).
Summing up, Eq.\,(\ref{2-6}) describes the corrections to CNT which arise
by replacing the assumption of a sharp solid-liquid interface with a more
realistic finite width, 
in the case of isotropic surface tension and Tolman length.

\section{Shape fluctuations: the interface Hamiltonian}
\setcounter{equation}{0}
\renewcommand{\theequation}{3.\arabic{equation}}

A real cluster may be spherical only on average. Far from being static,
clusters fluctuate widely away from their mean shape~\cite{Filion,Zykova}.
To describe fluctuations, we switch from a description in terms of the
crystallinity OP to another in which the cluster shape itself
rises to the role of fundamental variable. We begin by deriving a
coarse-grained, purely geometrical Hamiltonian for the
cluster surface directly from the microscopic free-energy functional
(\ref{2-2}), under the assumption of small deviations of the interface
from planarity. The outcome is a Canham-Helfrich (CH)
Hamiltonian~\cite{Canham,Helfrich}, containing spontaneous-curvature and
bending penalty terms in addition to interface tension.

For the present derivation, we build on Refs.\,\cite{Kogon,Kassner}.
Other attempts to derive an effective interface Hamiltonian from a mean-field
density functional are described in \cite{Napiorkowski,Segovia-Lopez}.
Let the cluster ``surface'' be depicted as a closed mathematical surface
$\Sigma$ embedded in three-dimensional space and let ${\bf R}(u,v)$ be the
parametrization (coordinate patch) of an infinitesimal piece of $\Sigma$.
We switch from 3D cartesian coordinates, ${\bf r}=(x,y,z)$, to new coordinates
$q_\alpha=(u,v,\zeta)$ (tangential and normal to $\Sigma$) by the
transformation
\be
{\bf r}={\bf R}(u,v)+\zeta\widehat{\bf n}(u,v)\,,
\label{3-1}
\ee
where
\be
\widehat{\bf n}(u,v)=
\frac{{\bf R}_u\wedge{\bf R}_v}{|{\bf R}_u\wedge{\bf R}_v|}
\label{3-2}
\ee
is the unit normal to $\Sigma$. For a patch that deviates only slightly from
planarity, we may adopt a free energy ${\cal G}[\phi_0(\zeta(x,y,z))]$,
thus arriving at the surface Hamiltonian
\be
{\cal H}_s[\Sigma]=\int{\rm d}u\,{\rm d}v\,{\rm d}\zeta\,J
\left\{\frac{c}{2}\left(\nabla\phi_0(\zeta)\right)^2+\frac{\kappa}{2}
\left(\nabla^2\phi_0(\zeta)\right)^2+g(\phi_0(\zeta))\right\}
\label{3-3}
\ee
with $J=|{\bf r}_u\cdot({\bf r}_v\wedge{\bf r}_\zeta)|=
|\widehat{\bf n}\cdot({\bf r}_u\wedge{\bf r}_v)|$. In order to make
Eq.\,(\ref{3-3}) simpler, it is convenient to view the patch as parametrized
in terms of orthonormal, arc-length coordinates, i.e.,
${\bf R}_u\cdot{\bf R}_v=0$ and $|{\bf R}_u|=|{\bf R}_v|=1$ all over the
patch. Although this construction is rigorously possible only for surfaces
having zero Gaussian curvature ($K=0$)~\cite{Abate},
we can reasonably expect that only small errors of order $K$ are made
for quasiplanar
interfaces. With this caution in mind, we go on to get (see Appendix B):
\ba
\frac{\partial{\bf r}}{\partial u} &=& (1-\zeta\kappa_n^{(1)}){\bf R}_u-
\zeta\tau_g{\bf R}_v\,;
\nonumber \\
\frac{\partial{\bf r}}{\partial v} &=& -\zeta\tau_g{\bf R}_u+
(1-\zeta\kappa_n^{(2)}){\bf R}_v\,;
\nonumber \\
\frac{\partial{\bf r}}{\partial\zeta} &=& \widehat{\bf n}\,,
\label{3-4}
\ea
where $\kappa_n^{(1)}$ and $\kappa_n^{(2)}$ are the normal curvatures of
the $u$- and $v$-lines respectively, and $\tau_g\equiv\tau_g^{(1)}=
-\tau_g^{(2)}$ is the geodetic torsion.
From Eqs.\,(\ref{3-4}), we readily derive the metric tensor
$g_{\alpha\beta}$,
\be
g_{\alpha\beta}\equiv\frac{\partial{\bf r}}{\partial q_\alpha}\cdot
\frac{\partial{\bf r}}{\partial q_\beta}
=\left(\begin{array}{ccc}
\left(1-\zeta\kappa_n^{(1)}\right)^2+\zeta^2\tau_g^2 & -2\zeta\tau_g+
\zeta^2\tau_g\left(\kappa_n^{(1)}+\kappa_n^{(2)}\right) & 0 \\
-2\zeta\tau_g+\zeta^2\tau_g\left(\kappa_n^{(1)}+\kappa_n^{(2)}\right)
& \left(1-\zeta\kappa_n^{(2)}\right)^2+\zeta^2\tau_g^2 & 0 \\
0 & 0 & 1
\end{array}\right)\,,
\label{3-5}
\ee
and the Jacobian,
\be
J=\left(1-\zeta\kappa_n^{(1)}\right)\left(1-\zeta\kappa_n^{(2)}\right)
-\zeta^2\tau_g^2=\sqrt{g}\,,
\label{3-6}
\ee
$g$ being the determinant of (\ref{3-5}).
Considering that covariant and contravariant components of a vector
are built by projecting it on the bases $\nabla q_\alpha$ and
$\partial{\bf r}/\partial q_\alpha$, respectively,
we can calculate the gradient of a scalar field $\phi$ and the divergence
of a vector field {\bf A} in local coordinates as follows:
\be
\nabla\phi=\frac{\partial\phi}{\partial q_\alpha}g^{\alpha\beta}
\frac{\partial{\bf r}}{\partial q_\beta}\,\,\,\,\,\,{\rm and}\,\,\,\,\,\,
\nabla\cdot{\bf A}=\frac{1}{\sqrt{g}}\frac{\partial}{\partial q_\alpha}
\left(\sqrt{g}A^\alpha\right)\,,
\label{3-7}
\ee
$g^{\alpha\beta}$ being the inverse of (\ref{3-5}). In particular,
\be
\nabla\phi(\zeta)=\phi^{\prime}(\zeta)\widehat{\bf n}\,\,\,\,\,\,{\rm and}
\,\,\,\,\,\,\nabla^2\phi(\zeta)=\phi^{\prime\prime}(\zeta)+\phi^{\prime}(\zeta)
\nabla\cdot\widehat{\bf n}\,,
\label{3-8}
\ee
where
\be
\nabla\cdot\widehat{\bf n}=\frac{1}{\sqrt{g}}
\left(-\kappa_n^{(1)}-\kappa_n^{(2)}-2\zeta\tau_g^2\right)\,.
\label{3-9}
\ee
Finally, the mean and Gaussian curvatures of the patch are given by
\be
H=\left.\frac{1}{2}\nabla\cdot\widehat{\bf n}\right|_{\zeta=0}=
-\frac{1}{2}\left(\kappa_n^{(1)}+\kappa_n^{(2)}\right)
\label{3-10}
\ee
and
\be
K=\widehat{\bf n}\cdot\left(\frac{\partial\widehat{\bf n}}
{\partial u}\wedge\frac{\partial\widehat{\bf n}}{\partial v}\right)=
\kappa_n^{(1)}\kappa_n^{(2)}-\tau_g^2\,.
\label{3-11}
\ee
Hence, 1) the mean curvature, which is defined only up to a sign depending
on our convention on the orientation of $\hat{\bf n}$, is half the sum of
the two normal curvatures relative to any orthogonal parametrization, i.e.,
not necessarily the two principal curvatures; 2) since $K$ is the product
of the two principal curvatures, the geodetic torsion must vanish when the
coordinate lines are also lines of curvature.

We are now in a position to simplify Eq.\,(\ref{3-3}).
Upon using Eq.\,(\ref{2-4}) to eliminate $g(\phi_0)$ in favor of
$(c/2)\phi_0^{\prime\,2}+(3\kappa/2)\phi_0^{\prime\prime\,2}-
\kappa\left(\phi_0^{\prime}\phi_0^{\prime\prime}\right)^{\prime}$,
and inserting Eqs.\,(\ref{3-6}), (\ref{3-8}), (\ref{3-10}), and
(\ref{3-11}), we eventually get
\ba
{\cal H}_s &=& \int{\rm d}u\,{\rm d}v\,{\rm d}\zeta\,
\left(1+2\zeta H +\zeta^2 K\right)\left\{c\phi_0^{\prime 2}(\zeta)+
\frac{3}{2}\kappa\phi_0^{\prime\prime 2}(\zeta)\right.
\nonumber \\
&+& \left.\frac{\kappa}{2}\left(\phi_0''(\zeta)+
\phi_0'(\zeta)\frac{2H-2\zeta\tau_g^2}{1+2\zeta H +\zeta^2 K}
\right)^2-\kappa\left(\phi_0'(\zeta)\phi_0''(\zeta)\right)'\right\}\,.
\label{3-12}
\ea
We now argue that, to a first approximation, any term of order higher than
$H^2$ and $K$ can be discarded. Moreover, $\int{\rm d}u{\rm d}v=\int{\rm d}S$
since $|{\bf R}_u\wedge{\bf R}_v|=1$. Lastly, the geodetic torsion vanishes
if we perform a change of integration variables (that is,
a change of parametrization) such that the coordinate lines are also lines of
curvature~\cite{Abate2}.
In the end, we are left with the classic Canham-Helfrich Hamiltonian for
fluid membranes:
\be
{\cal H}_s=\int_\Sigma{\rm d}S\,\left(\overline{a}+\overline{b}H+
\overline{c}H^2+\overline{d}K\right)\,,
\label{3-13}
\ee
with the following explicit expressions for the coefficients:
\ba
\overline{a} &=& \int_{-\infty}^{+\infty}{\rm d}\zeta
\left[c\phi_0^{\prime\,2}(\zeta)+2\kappa
\phi_0^{\prime\prime\,2}(\zeta)\right]\,;
\nonumber \\
\overline{b} &=& 2\int_{-\infty}^{+\infty}{\rm d}\zeta\,\zeta
\left[c\phi_0^{\prime\,2}(\zeta)+2\kappa
\phi_0^{\prime\prime\,2}(\zeta)\right]\,;
\nonumber \\
\overline{c} &=& 2\kappa\int_{-\infty}^{+\infty}{\rm d}\zeta\,
\phi_0^{\prime\,2}(\zeta)\,;
\nonumber \\
\overline{d} &=& \int_{-\infty}^{+\infty}{\rm d}\zeta\left\{\zeta^2
\left[c\phi_0^{\prime\,2}(\zeta)+2\kappa
\phi_0^{\prime\prime\,2}(\zeta)\right]-\kappa
\phi_0^{\prime\,2}(\zeta)\right\}\,.
\label{3-14}
\ea

A few remarks are now in order:
1) $H$ and $K$ are reparametrization invariants,
hence no ambiguity arises from the arbitrariness of the parametrization used.
2) The above derivation actually applies for just one
$\Sigma$ patch. However, upon viewing $\Sigma$ as the union of many disjoint
patches, the Hamiltonian (\ref{3-13}) holds for the whole $\Sigma$ as well.
3) As anticipated, the coefficient $\overline{d}$ of the
$K$ term in (\ref{3-13}) could be different from the quoted one since a
parametrization in terms of orthonormal coordinates does not generally
exist. However, as far as we only allow for clusters with the topology
of a sphere, $\int_\Sigma{\rm d}S\,K$ takes the constant value of $4\pi$
by the Gauss-Bonnet theorem and the $K$ term in ${\cal H}_s$ can be dropped.
Upon comparing the definition of $\overline{a}$ and $\overline{b}$ in
Eqs.\,(\ref{3-14}) with Eqs.\,(\ref{2-8}), we can rewrite Eq.\,(\ref{3-13})
in the form (restoring everywhere the dependence upon interface orientation):
\be
{\cal H}_s=\int_\Sigma{\rm d}S\,\left(\sigma_m(\hat{\bf n})-
2\sigma_m(\hat{\bf n})\delta_m(\hat{\bf n})H+2\lambda(\hat{\bf n})H^2\right)\,,
\label{3-15}
\ee
where $\lambda=\overline{c}/2$ (we note that
$\lambda=\kappa\phi_{s0}^2/(3\ell)$
under the same hypotheses for which Eq.\,(\ref{a-15}) holds).
4) The term linear in $H$ is related to the {\em spontaneous
curvature} of
$\Sigma$, $H_0=-\overline{b}/(2\overline{c})$, which is proportional to the
Tolman length $\delta_m$.
A nonzero value of $H_0$ yields a difference in energy
between inward and outward interface protrusions, thus entailing a non-zero
$\delta_m$. The additional fact that in systems, such as the Ising model,
where the symmetry is perfect between the two phases then $\delta_m=0$,
has long been known~\cite{Fisher}.

We point out that Eq.\,(\ref{3-15}) retains the same form as
in the isotropic case~\cite{Prestipino}. In the general anisotropic case,
the dependence of the Hamiltonian parameters on the interface normal is
through the constants $c$ and $\kappa$, and the function $\phi_0(z)$.

\section{The cluster free energy in two extreme cases: \\
isotropic and strongly anisotropic interface tension}
\setcounter{equation}{0}
\renewcommand{\theequation}{4.\arabic{equation}}

Considering that every single realization of the profile of the cluster
surface should be sampled in equilibrium with a weight proportional to
$\exp\{-\beta{\cal H}_s\}$, it is natural to define a volume-dependent cost
of cluster formation through
\be
\Delta G(V)=-\rho_s|\Delta\mu|V+F_s(V)
\label{4-1}
\ee
with
\be
F_s(V)=-k_BT\ln Z_s(V)=-k_BT\ln\left\{a^3\int{\cal D}\Sigma\,
e^{-\beta{\cal H}_s}\,\delta({\cal V}[\Sigma]-V)\right\}\,.
\label{4-2}
\ee
In the above expression of the constrained partition function $Z_s$,
$a=\rho_s^{-1/3}$ is a microscopic length of the system,
${\cal V}[\Sigma]$ is the volume enclosed by the closed surface $\Sigma$, and
${\cal D}\Sigma$ a yet-to-be-specified integral measure.

While the calculation of $F_s$ for a realistic form of
$\hat{\bf n}$-dependent parameters in (\ref{3-15}) is certainly
possible numerically once the admissible surfaces have been parametrized
in terms of a basis of eigenfunctions,
some restrictions are to be made in practice if we want to make
analytical progress.
In the following, we examine two limiting cases for $\sigma_m(\hat{\bf n})$,
according to whether it is constant or strongly anisotropic.
In general, a strongly anisotropic $\sigma_m$ is typical of e.g. systems
where melting is very strongly first order, implying very sharp and
thus direction dependent solid-liquid interfaces, such as for example
in the case of alkali halides~\cite{Zykova2}. That brings about a non-spherical
cluster shape through the prescription that the surface free energy be the
minimum possible for the given cluster volume $V$. The same condition
is responsible for a spherical shape when the interface free energy is
isotropic.

\subsection{Isotropic interfaces}

If $\sigma_m,\delta_m$, and $\lambda$ in Eq.\,(\ref{3-15}) do not depend on
$\hat{\bf n}$, the shape of a cluster is on average spherical.
We here compute the free energy (\ref{4-2}) assuming small
deviations from this shape.

Neglecting overhangs and liquid inclusions, let $r=R(\theta,\phi)$ be the
equation of $\Sigma$ in spherical coordinates.
We assume only small deviations from a sphere,
i.e., $R(\theta,\phi)=R_0[1+\epsilon(\theta,\phi)]$, with
$\epsilon(\theta,\phi)\ll 1$~\cite{Milner}.
Then, we expand $\epsilon(\theta,\phi)$ in real spherical harmonics,
\be
\epsilon(\theta,\phi)=
\sum_{l=1}^\infty\sum_{m=-l}^lx_{l,m}Y_{l,m}(\theta,\phi)\,,
\label{4-3}
\ee
and we agree to ignore, from now on, all terms beyond second-order in the
coefficients $x_{l,m}$. With these specifications,
we obtain approximate expressions for the area of $\Sigma$ and its
enclosed volume, as well as for the mean curvature $H$.
Upon inserting this form of ${\cal H}_s$ in terms of the $x_{l,m}$
into Eq.\,(\ref{4-2}), we are left with the evaluation of a Gaussian integral.
While we refer the reader to Appendix C for all the technicalities,
we here quote the result of the calculation. The free energy cost of
cluster formation for large $V$ is
\ba
\Delta G(V)&=&-\rho_s|\Delta\mu|V+(36\pi)^{1/3}\sigma^{\rm QS}V^{2/3}-
(384\pi^2)^{1/3}\sigma^{\rm QS}\delta^{\rm QS}V^{1/3}
\nonumber \\
&+&4\pi\sigma^{\rm QS}\epsilon^{\rm QS}-\frac{7}{6}k_BT\ln\left((36\pi)^{1/3}
\left(\frac{V}{a^3}\right)^{2/3}\right)\,,
\label{4-4}
\ea
where $\sigma^{\rm QS},\delta^{\rm QS}$, and $\epsilon^{\rm QS}$
can be read in Eq.\,(\ref{c-21}).
The above formula is strictly valid only near coexistence, where the
various assumptions beneath its derivation are expected to hold true.
We have thus found that the surface free energy has a form consistent
with the Dillmann-Meier ansatz, with $T$-dependent parameters
$\sigma^{\rm QS},\delta^{\rm QS}$, and $\epsilon^{\rm QS}$ that are
different (even at $T_m$!) from the corresponding ones in Landau-theory
$\sigma^{\rm L},\delta^{\rm L}$, and $\epsilon^{\rm L}$, and with a universal
logarithmic correction to the mean-field form of $\Delta G$. This term
is responsible for the well known $R^{*7/3}$ exponential prefactor to the
nucleation rate~\cite{Guenther}.

\subsection{Anisotropic interfaces}

We now consider an interface tension of the form:
\be
\sigma(\hat{\bf n})=\sigma_{100}\left[1+
M\left(\hat{n}_x^4+\hat{n}_y^4+\hat{n}_z^4-1\right)^2\right]
\label{4-5}
\ee
with $M\rightarrow\infty$, written in terms of the cartesian components of
the outer normal to the cluster surface. In the infinite-$M$ limit, the
equilibrium crystal shape is a cube, though rectangular cuboids are also
admissible, though not optimal, shapes (they arise at non-zero temperatures).
The terms in Eq.\,(\ref{3-15}) beyond the first are singular in the
$M\rightarrow\infty$ limit; however, they would contribute to the surface
free energy if $M$ were large but not infinite, see more in Appendix D.
In the same Appendix we show that the asymptotic, large-$V$ free-energy
cost of cluster formation is given by:
\be
\Delta G(V)=-\rho_s|\Delta\mu|V+6\sigma^{PP}V^{2/3}+12\nu^{PP}V^{1/3}+
k_BT\ln\left(6\left(\frac{V}{a^3}\right)^{2/3}\right)+{\rm const.}
\label{4-6}
\ee
with $\sigma^{PP}\equiv\sigma_{100}$. Similarly to the isotropic case,
in the cluster free energy (\ref{4-6}) both a logarithmic term and an
offset are added to the classical CNT expression of $\Delta G$ for a
cubic cluster of side $V^{1/3}$.
The Tolman term in Eq.\,(\ref{4-6}) only appears if we envisage an energy
penalty, that is $\nu^{PP}$ per unit length, also for the edges.

More generally, in all the anisotropic-nucleation models examined in
Appendix D, the consideration of clusters of same type but unequal
edges/semiaxes provides for ``breathing'' fluctuations of the surface
that determine the appearance of a logarithmic term in $\Delta G$.
In fact, for all such models, the analytically computed $\Delta G(V)$ is
asymptotically given, as in Eq.\,(\ref{4-6}), by the CNT expression --
as written for the respective symmetric shape -- plus subleading
terms in the form of a Tolman term, a universal logarithm
($ck_BT\ln V^{(d-1)/d}$ in $d$ dimensions), and a negative offset.
The value of $c$ is $1/2$ for rectangles and 1 for both cuboids and
ellipsoids. This is to be contrasted with the quasispherical-cluster case,
where $c=-7/6$ by Eq.\,(\ref{4-4}). Apparently, the value of $c$ is
sensitive to both the space dimensionality and the number of independent
parameters that are needed to describe the cluster shape, in turn crucial
to determine the entropy contents of the surface degrees of freedom
(for a quasispherical cluster, this number of parameters goes to infinity
with $V$). 
In short, a large anisotropy in the interface tension has the overall effect of
drastically reducing the spectrum of thermal fluctuations of cluster shape.
The reduction cancels the entropy gain which these fluctuations produced 
in the isotropic case.

This attractive prediction is a difficult one to fully validate numerically
at present. A logarithmic correction to CNT can only be detected if we
push the numerical investigation of $\Delta G(V)$ so close to
coexistence as to make the Dillmann-Meier form exact for all but the smallest
clusters, and that is still a difficult task (see more in the next Section).
In the near future, with faster computers becoming available, we can imagine
that it will be possible to directly probe the cluster geometry through the
optimization of the logarithmic prefactor in an ansatz of the kind
(\ref{4-4}) or (\ref{4-6}), and thus choose among the many cluster models
on the market the one which is most appropriate to the problem at hand.

\section{Numerical assessment of the theory}
\setcounter{equation}{0}
\renewcommand{\theequation}{5.\arabic{equation}}

We now critically consider if there are signatures of the degree of
anisotropy of the interface free energy in the free-energy cost of
cluster formation for a specific instance of microscopic interaction.

We first recall how the work of formation of a $n$-particle cluster is
calculated from simulationsi~\cite{tenWolde,Reiss,Bowles}.
Given a criterion to identify solid-like clusters
within a predominantly liquid system of $N$ particles,
the average number of $n$-clusters is given, for $1\ll n\simeq n^*$,
by $N_n=N\,e^{-\beta(G_n-n\mu_l)}$, where $\mu_l$ is the chemical potential
of the liquid and $G_n$ is the ${\cal O}(n)$ Gibbs free
energy of the $n$-cluster, including also the
contribution associated with the wiggling of the cluster center of mass
within a cavity of volume $V/N$ (observe that CNT estimates
$G_n$ as $n\mu_s+c\sigma(n/\rho_s)^{2/3}$, where $\mu_s$ is the chemical
potential of the solid and $c$ a geometrical factor). For rare clusters,
it thus follows that $\Delta G(n)\equiv G_n-n\mu_l=-k_BT\ln(N_n/N)$.
This equation is then taken to represent the work of cluster formation for
all $n>1$. Maibaum~\cite{Maibaum} has shown that the same formula applies
for the Ising model.

However, for quenches that are not too deep, the spontaneous occurrence of
a large solid cluster in the metastable liquid is a rare event. This poses
a problem of poor statistics in the Monte Carlo (MC) estimation of $N_n$,
which is overcome through e.g. the use of a biasing potential that couples
with the size $n_{\rm max}$ of the largest cluster. In practice, this keeps
the system in the metastable state for all the $n$'s of interest.
By properly reweighting the sampled microstates one eventually recovers the
ordinary ensemble averages.
This umbrella-sampling (US) method was used in Refs.~\cite{tenWolde,Pan}
to compute $\Delta G(n)$ for the Lennard-Jones fluid and the 3D Ising model,
respectively. The main obstacle to the calculation of $\Delta G(n)$ by US
is the necessity of performing the identification of the largest cluster
in the system after every MC move. This problem can be somewhat mitigated
by the use of a hybrid MC algorithm~\cite{Gelb}, which in our case reduced
the simulation time by a factor of about 20.

A low-temperature Ising magnet where the majority of spins point against
the applied field probably yields the simplest possible setup for the
study of nucleation. Along the first-order transition line of the model,
where two (``up'' and ``down'') ferromagnetic phases coexist, the interface
(say, (100)) between the two phases undergoes a roughening transition
at a certain $T=T_R$. The up-down interface tension at coexistence is
strongly anisotropic close to zero temperature; moreover, it is either
singular or smooth according to whether $T$ is below or above $T_R$.
Strictly speaking, the interface tension is anisotropic also above $T_R$,
though less and less so when approaching the critical temperature $T_c$
from below~\cite{Mon,Hasenbusch1}.
Exactly at $T_c$ the interface tension critically vanishes~\cite{Rottman}.
When a sample originally prepared in the ``down'' phase is slightly
pushed away from coexistence by a small positive field and thus made
metastable, the critical droplet of the ``up'' phase is expected to be
less and less spherical as $T$ decreases.

With the 3D Ising model as a test system, we carried out a series of
extensive US simulations, computing the cluster free energy $\Delta G(n)$
relative to the nucleation process of magnetization reversal for a fixed
$T=0.6\,T_c$, slightly above the roughening temperature $T_R$ of the (100)
facet ($T_R=0.5438\ldots\,T_c$~\cite{Hasenbusch2}), and for a number of
values of the external field $h$ ($0.30,0.35,\ldots,0.65$, in $J$ units).
Two up spins are said to belong to the same cluster if there is a sequence
of neighboring up spins between them; the counting of clusters was done with
the Hoshen-Kopelman algorithm~\cite{Hoshen}.
The absolute value of $\Delta G(1)$ was determined through a standard MC
simulation of the system with all spins down, with no bias imposed on the
sampling of the equilibrium distribution.
We point out that, at the chosen temperature, the Ising
surface tension is barely anisotropic~\cite{Mon}, which would exclude a net
preference for either the spherical or the cubic shape. Furthermore, we are
sufficiently far away from $T_c$ not to worry about the percolation
transition of geometric clusters which was first described in
\cite{MuellerKrumbhaar}. This event, which would invalidate the
assumption (at the heart of the conventional picture of nucleation) of a
dilute gas of clusters, is still far away here.

%
%   FIGURE 1
%
\begin{figure}
\includegraphics[width=10cm]{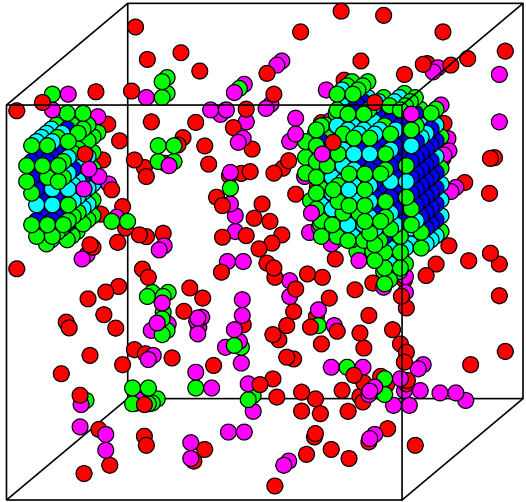}
\caption{(Color online). A snapshot taken from our Monte Carlo simulation
of the 3D Ising model at $T=0.6\,T_c$ and $h=0.30$, showing a cluster of
$n=685$ up spins, i.e., close to the critical size for that $h$. Up spins
are differently colored according to the number of nearest-neighboring up
spins (blue, 6; cyan, 5; green, 2-4; magenta, 1; red, 0). Down spins are
not shown.}
\label{fig1}
\end{figure}

Coherently with the physical picture at the basis of our theory, we verified
for all the $h$ considered that clusters close to critical indeed contain
the vast majority of up spins in the system.
A sample of the critical cluster for $h=0.30$ is shown in Fig.\,1.
Looking at this picture, it is hard to say whether this particular
realization of the critical cluster resembles more a sphere or a cube.
When moving to $h=0$, a spherical shape is eventually preferred
over the cube far above $T_R$, whereas the opposite occurs much below $T_R$.

%
%   FIGURE 2
%
\begin{figure}
\includegraphics[width=14cm]{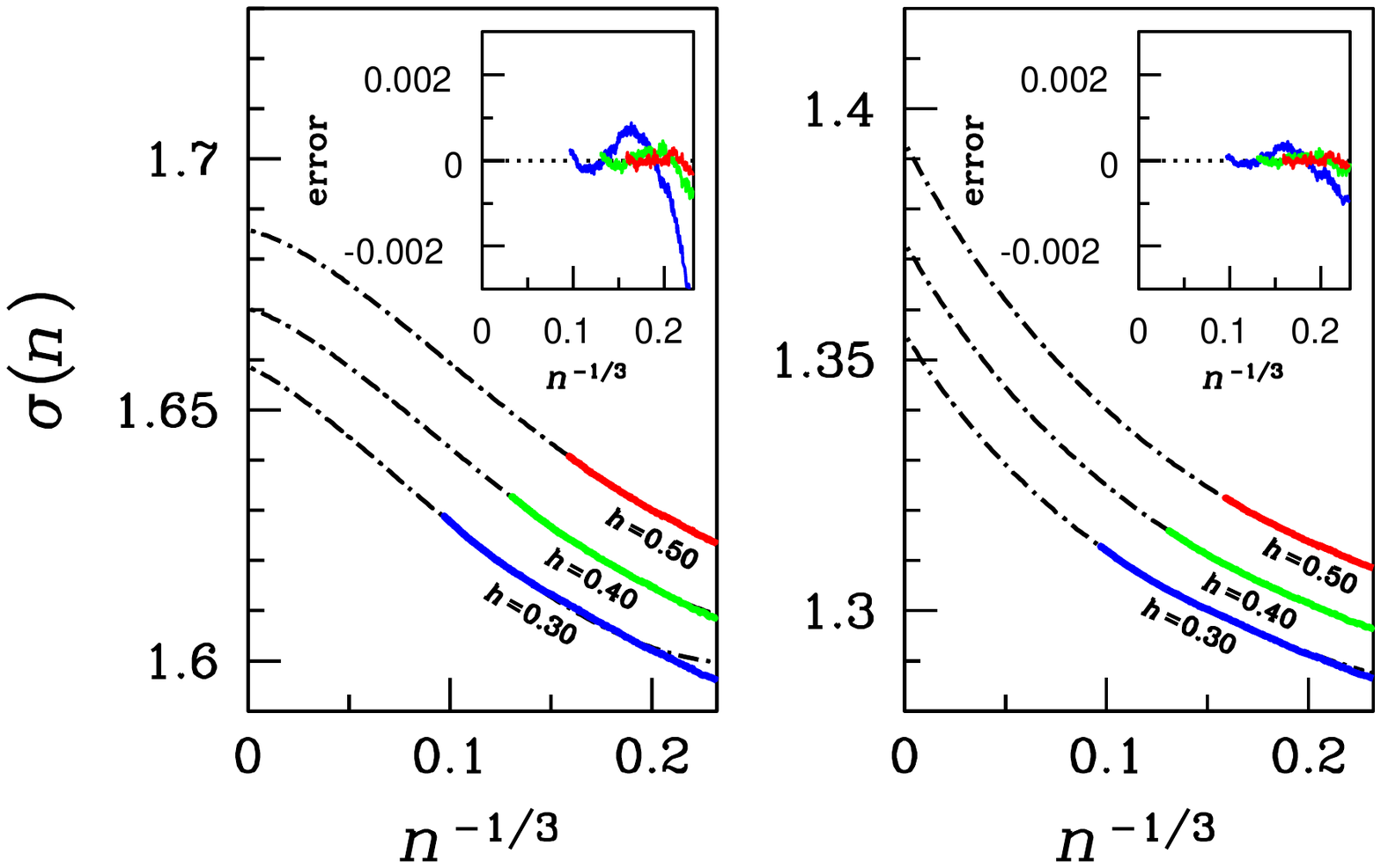}
\caption{(Color online).
The cluster free energy $\sigma^{\rm I}$ of the 3D Ising model on a
cubic lattice in units of $J/a^2$ is plotted as a function of $n^{-1/3}$
(and up to $80^{-1/3}$)
for three values of $h$ and for $T=0.6\,T_c$ ($a$ is the lattice
spacing and $J>0$ is the spin-coupling constant). The lattice includes
$20^3$ sites ($25^3$ for $h=0.30$).
Umbrella-sampling simulations consisted of 4M equilibrium sweeps
for each $n$ window (one window covering eleven values of $n$).
Thick colored lines, MC data; black lines, least-square fits of the $n>80$
data points for $h=0.30,0.40,0.50$, based on Eq.\,(\ref{4-4}) (left)
and (\ref{4-6}) (right). Data plotted in the two panels
look different simply because the expressions of cluster area $S(n)$ are
different between left and right (see text).
Inset, the difference between the raw data and the fit.}
\label{fig2}
\end{figure}

In Fig.\,2, the ratio $\sigma^{\rm I}(n)$ of the surface free energy
$F_s(n)=\Delta G(n)+|\Delta\mu|n$ to the area $S(n)$ of
the cluster surface is reported as a function of $n^{-1/3}$, and the data
are fit using the functions (\ref{4-4}) and (\ref{4-6}) (we stress that
different expressions apply for $S(n)$
on the left and right panels of Fig.\,2, i.e., $(36\pi)^{1/3}n^{2/3}$ and
$6n^{2/3}$ respectively; accordingly, the spherical $\sigma$'s
would typically turn out a factor
$6/(36\pi)^{1/3}$ larger than the cubic $\sigma$'s).
Both fits are based on three parameters, namely  $\sigma$, $\delta$,
and $\epsilon$, which enter in a different way in Eqs.\,(\ref{4-4}) and
(\ref{4-6}). However, the dependence on $n$ is similar for the
two fitting functions, except for the numerical factor in front of the
(parameter-free) $n^{-2/3}\ln n$ term. Looking at Fig.\,2, it appears
that the quality of the ``cubic'' fit is slightly better than that of the
``spherical'' fit, in line with the fact that, for
$T\gtrsim T_R$, the Ising surface tension is moderately anisotropic.
Clearly, at $T=0.6\,T_c$ the nucleus is neither spherical nor cubic, and
one may object that neither of the fits would actually be meaningful.
We nonetheless argue that, within the uncertainty associated with the
finite $h$ value in the simulations, the better one of the fits will
correspond to the regular shape which is closest to that of the real
nucleus, thus giving a qualitative indication of the prevailing isotropic
or anisotropic character of the solid-liquid interface tension.
When going to smaller and smaller $h$, and provided $T$
is sufficiently above $T_R$, we expect that the ``spherical''
fit would eventually become better than the ``cubic'' fit.

\section{Conclusions}

In order to estimate from nucleation the solid-liquid interface free energy
$\sigma_m$ of a substance, two indirect routes are available: one is through
the measurement of the solid nucleation rate as a function of
temperature (see e.g. \cite{Li,Li2,Franke}),
the other is via the free energy $\Delta G$ of solid-cluster formation in a
supercooled-liquid host, as determined for example in a numerical simulation
experiment for a system model. In both cases, the theoretical framework of
classical nucleation theory (CNT) has routinely been employed to
extract $\sigma_m$.
This is far from satisfactory, as discussed at length in
Ref.\,\cite{Prestipino} and in many other papers, due to the neglected
cluster interface-tension dependence on both the droplet volume $V$ and the
supersaturation $|\Delta\mu|$.

Concentrating on the expression of the cluster formation energy $\Delta G$
as a function of $V$ and $\Delta\mu$, we gave here an extension of the modified
CNT theory first introduced in \cite{Prestipino}, now including anisotropy,
which is important when only a few interface orientations survive in the
equilibrium average cluster shape.
We showed that, also in this case, a universal non-CNT $\ln V$ term
is found in the asymptotic expression of the surface free energy versus
volume, so long as an infinity of regular shapes is allowed to occur.
However that term has now a different prefactor with respect to
the quasispherical case. In particular, the sign is positive
for large anisotropy and negative for vanishing anisotropy.
The sign of that prefactor, which we surmise is related
to the amount of surface entropy developed by cluster shape fluctuations,
is proposed as the imprinted signature of the geometrical shapes most
preferred by the nucleation cluster -- negative for spherical or
very isotropic shapes, positive for nearly polyhedric or anyway very
anisotropic shapes. For the 3D Ising model slightly above the (100) roughening
transition temperature, the detected sign suggests cubic rather than
spherical cluster symmetry for moderate supersaturation/external field.
Much more work and larger simulation sizes should be needed in the future
in order to verify the expected change of sign of the $\ln V$ term as
spherical shapes will be approached closer and closer to the coexistence
line when the temperature is quite larger than $T_R$
(though still far from the critical region).

\begin{center}
Acknowledgements
\end{center}

This project was
co-sponsored by CNR through ESF Eurocore Project FANAS AFRI, by the
Italian Ministry of Education and Research through PRIN COFIN Contract
2010LLKJBX004, by SNF Sinergia Project CRSII2\_136287/1, and by EU
ERC Advanced Grant 320796.

\appendix
\section{Calculation of $\sigma_m$ and $\delta_m$}
\setcounter{equation}{0}
\renewcommand{\theequation}{A.\arabic{equation}}

In this Appendix, we provide approximate expressions for the quantities
$\sigma_m$ and $\delta_m$ in Eqs.\,(\ref{2-8}) for a specific model of
homogeneous-system free energy $g(\phi)$ in the functional (\ref{2-2}).

Once the exact OP profile $\phi_0(z)$ of the planar interface
has been determined for
the given $g$, the explicit values of $\sigma_m$ and $\delta_m$, and of
$\epsilon_m\equiv\epsilon^{\rm L}(T_m)$ can be computed. While $\sigma_m$
and $\epsilon_m$ are strictly positive quantities, the sign of $\delta_m$
is not {\em a priori} definite. A special but sufficiently general case
of $g$ function is the following:
\be
g(\phi;T=T_m)=c_{20}\phi^2\left(1-\frac{\phi}{\phi_{s0}}\right)^2
\left[1+(\gamma_5+2\gamma_6)\frac{\phi}{\phi_{s0}}+\gamma_6
\frac{\phi^2}{\phi_{s0}^2}\right]
\label{a-1}
\ee
with $\gamma_5>-1-3\gamma_6$ for $0<\gamma_6\le 1$ and $\gamma_5>-2\gamma_6-
2\sqrt{\gamma_6}$ for $\gamma_6>1$.
Equation (\ref{a-1}) is the most general sixth-degree polynomial which
admits two non-equivalent minimum valleys at 0 and $\phi_{s0}$, and no
further negative minimum between them.
For this $g$, the differential equation (\ref{2-4}) is still too difficult
to solve in closed form for generic $\kappa$, even when $\gamma_5=\gamma_6=0$.
Hence, we decided to work perturbatively in $\kappa,\gamma_5$, and $\gamma_6$.

At zeroth order, i.e., $\kappa=\gamma_5=\gamma_6=0$, corresponding to
$\phi^4$ theory, the solution to (\ref{2-4}) is
\be
\overline{\phi}_0(z)=\frac{\phi_{s0}}{2}
\left\{1-\tanh\left(\frac{z-C}{\ell}\right)\right\}
\label{a-2}
\ee
with $\ell=\sqrt{2c/c_{20}}$ and arbitrary $C$. We fix $C$ by requiring that
the interface is centered at $z=0$ (hence $C=0$). Then, by still keeping
$\gamma_5=\gamma_6=0$, we switch on $\kappa$ and search for a second-order
solution to Eq.\,(\ref{2-4}) in the form
\be
\phi_0(z)=\overline{\phi}_0(z)+\frac{\kappa}{c\ell^2}\chi_1(z)+
\left(\frac{\kappa}{c\ell^2}\right)^2\chi_2(z)\,.
\label{a-3}
\ee
We thus arrive at the two equations:
\be
c\overline{\phi}_0'\chi_1'-g_0'(\overline{\phi}_0)\chi_1=c\ell^2\left(
\overline{\phi}_0'\overline{\phi}_0'''-\frac{1}{2}\overline{\phi}_0''^2\right)
\label{a-4}
\ee
and
\be
c\overline{\phi}_0'\chi_2'-g_0'(\overline{\phi}_0)\chi_2=
c\ell^2\left(\overline{\phi}_0'\chi_1'''+
\chi_1'\overline{\phi}_0'''-\overline{\phi}_0''\chi_1''\right)-
\frac{c}{2}\chi_1'^2+\frac{g_0''(\overline{\phi}_0)}{2}\chi_1^2\,,
\label{a-5}
\ee
where
\be
g_0(\phi)=c_{20}\phi^2\left(1-\frac{\phi}{\phi_{s0}}\right)^2\,.
\label{a-6}
\ee
By requiring that $\phi_0(z)$ is centered at $z=0$ we obtain
\be
\chi_1(z)=\frac{\phi_{s0}}{\cosh^2(z/\ell)}
\left(2\tanh\frac{z}{\ell}-\frac{z}{\ell}\right)
\label{a-7}
\ee
and
\be 
\chi_2(z)=\frac{\phi_{s0}}{\cosh^2(z/\ell)}\left( 
32\tanh^3\frac{z}{\ell}-12\frac{z}{\ell}\tanh^2\frac{z}{\ell}-
8\tanh\frac{z}{\ell}+2\left(\frac{z}{\ell}\right)^2\tanh\frac{z}{\ell}-
3\frac{z}{\ell}\right)\,.
\label{a-8}
\ee
Hence, we find $\delta_m=0$ since the function
$c\phi_0'^2(z)+2\kappa\phi_0''^2(z)$ is even. Actually, the result
$\delta_m=0$ is valid at any order in $\kappa$ when $\gamma_5=\gamma_6=0$
(see below). Up to second order in $\kappa$, the values of $\sigma_m$ and
$\epsilon_m$ are given by:
\ba
\sigma_m&=&\left[1+\frac{2}{5}\,\frac{\kappa}{c\ell^2}-\frac{38}{35}\,
\left(\frac{\kappa}{c\ell^2}\right)^2\right]\frac{c\phi_{s0}^2}{3\ell}\,;
\nonumber \\
\epsilon_m&=&\left[\frac{\pi^2-6}{12}+
\left(\frac{26}{5}-\frac{\pi^2}{3}\right)\frac{\kappa}{c\ell^2}+
\left(\frac{1566}{175}-\frac{4\pi^2}{3}\right)
\left(\frac{\kappa}{c\ell^2}\right)^2\right]\ell^2\,.
\label{a-9}
\ea

Next, we take $\kappa,\gamma_5$, and $\gamma_6$ all non-zero and of the same
order of magnitude, and search for a first-order solution to (\ref{2-4}) in
the form
\be
\phi_0(z)=\overline{\phi}_0(z)+\gamma_5\psi_1(z)+\gamma_6\xi_1(z)+
\frac{\kappa}{c\ell^2}\chi_1(z)\,.
\label{a-10}
\ee
Upon inserting (\ref{a-10}) into Eq.\,(\ref{2-4}), we obtain two
independent equations for $\psi_1(z)$ and $\xi_1(z)$, namely
\be
c\overline{\phi}_0'\psi_1'-g_0'(\overline{\phi}_0)\psi_1=
\frac{\overline{\phi}_0g_0(\overline{\phi}_0)}{\phi_{s0}}
\label{a-11}
\ee
and
\be
c\overline{\phi}_0'\xi_1'-g_0'(\overline{\phi}_0)\xi_1=
\left(2\frac{\overline{\phi}_0}{\phi_{s0}}+
\frac{\overline{\phi}_0^2}{\phi_{s0}^2}\right)g_0(\overline{\phi}_0)\,,
\label{a-12}
\ee
while $\chi_1(z)$ is still given by Eq.\,(\ref{a-7}).
The solutions to Eqs.\,(\ref{a-11}) and (\ref{a-12}) such that each term of
(\ref{a-10}) separately meets the requirement of being centered at zero
are the following:
\be
\psi_1(z)=-\frac{\phi_{s0}}{8\cosh^2(z/\ell)}\left(1-\ln 2+
\frac{z}{\ell}-\ln\cosh\frac{z}{\ell}\right)
\label{a-13}
\ee
and
\be
\xi_1(z)=-\frac{\phi_{s0}}{8\cosh^2(z/\ell)}\left[3(1-\ln 2)+
3\frac{z}{\ell}-3\ln\cosh\frac{z}{\ell}-
\frac{1}{2}\tanh\frac{z}{\ell}\right]\,.
\label{a-14}
\ee
Upon plugging the by now specified $\phi_0(z)$ in the integrals defining
$\sigma_m,\delta_m$, and $\epsilon_m$, we eventually obtain the formulae:
\ba
\sigma_m&=&\left(1+\frac{1}{4}\gamma_5+\frac{13}{20}\gamma_6+
\frac{2}{5}\,\frac{\kappa}{c\ell^2}\right)\frac{c\phi_{s0}^2}{3\ell}\,,\,\,
\delta_m=\frac{5}{48}\left(\gamma_5+3\gamma_6\right)\ell\,,\,\,\,{\rm and}
\nonumber \\
\epsilon_m&=&\left[\frac{\pi^2-6}{12}-\frac{\pi^2-6}{48}\gamma_5-
\left(\frac{17\pi^2}{240}-\frac{1}{2}\right)\gamma_6+
\left(\frac{26}{5}-\frac{\pi^2}{3}\right)\frac{\kappa}{c\ell^2}\right]\ell^2\,.
\label{a-15}
\ea
We thus see that $\delta_m$ is generically non-zero and may be of both signs.

In conclusion, we give a proof that $\delta_m$ vanishes identically for
\be
g(\phi)=c_{20}\phi^2\left(1-\frac{\phi}{\phi_{s0}}\right)^2\,,
\label{a-16}
\ee
whatever $\kappa$ is (a different argument can be found in \cite{Fisher}).
Let $\phi(z)$ be a solution to Eq.\,(\ref{2-4}) obeying the boundary
conditions
\be
\phi(-\infty)=\phi_{s0}\,,\,\,\,\phi(+\infty)=0\,,\,\,\,
\phi'(\pm\infty)=\phi''(\pm\infty)=\ldots=0\,.
\label{a-17}
\ee
There is an infinite number of such solutions, differing from each other by
a simple translation. Let us first prove that
$\widetilde{\phi}(z)\equiv\phi_{s0}-\phi(-z)$ is also a solution to
(\ref{2-4}). We have:
\be
g(\widetilde{\phi}(z))=g(\phi(-z))\,;\,\,\,\widetilde{\phi}'(z)=\phi'(-z)\,;
\,\,\,\widetilde{\phi}''(z)=-\phi''(-z)\,;
\,\,\,\widetilde{\phi}'''(z)=\phi'''(-z)\,.
\label{a-18}
\ee
We thus see that
\ba
&& \kappa\widetilde{\phi}'(z)\widetilde{\phi}'''(z)-
\frac{c}{2}\widetilde{\phi}'^2(z)-\frac{\kappa}{2}\widetilde{\phi}''^2(z)+
g(\widetilde{\phi}(z))=
\nonumber \\
&& \kappa\phi'(-z)\phi'''(-z)-\frac{c}{2}\phi'^2(-z)-
\frac{\kappa}{2}\phi''^2(-z)+g(\phi(-z))=0\,,
\label{a-19}
\ea
since Eq.\,(\ref{2-4}) is satisfied by $\phi$ for any $z$.
Hence, $\widetilde{\phi}(z)$ obeys the differential equation (\ref{2-4}).
Moreover, like $\phi(z)$, $\widetilde{\phi}(z)$ also satisfies the conditions
(\ref{a-17}). This is not yet sufficient to conclude that $\widetilde{\phi}(z)$
and $\phi(z)$ are the same function since they could differ by a translation
along $z$. However, if among the infinite possibilities the one is selected
such that $\phi(0)=\phi_{s0}/2$, then $\widetilde{\phi}(0)=\phi_{s0}/2$ and
the two functions coincide: $\widetilde{\phi}(z)=\phi(z)$, implying
\be
\phi(z)+\phi(-z)=\phi_{s0}\,\,\,\,\,\,{\rm for\,\,any}\,\,z\,.
\label{a-20}
\ee

Upon differentiating (\ref{a-20}) with respect to $z$ we find that
$\phi'(-z)=\phi'(z)$ and the function $\phi'(z)$ is even.
This is enough to conclude that $\int{\rm d}z\,z\phi'(z)=0$ (the interface
is centered in 0). Differentiating (\ref{a-20}) once more, we obtain
$\phi''(-z)=-\phi''(z)$ and $\phi''(z)$ is an odd function of $z$
(while $\phi''^2(z)$ is even). As a result,
$\sigma_m\delta_m=-\int{\rm d}z\,z[c\phi'^2(z)+2\kappa\phi''^2(z)]=0$
and the proof is complete.

\section{Derivation of Eq.\,(\ref{3-4})}
\setcounter{equation}{0}
\renewcommand{\theequation}{B.\arabic{equation}}

Let $\ell$ be a curve in $\Sigma$ parametrized by the arc length $s$ and
denote $({\bf t},{\bf n},{\bf b})$ the Frenet trihedron in
${\bf R}(u_0,v_0)\in\ell$. Note that we are using nearly the same symbol for
the normal to $\Sigma$ ($\hat{\bf n}$) and for the normal vector to $\ell$ in
${\bf R}(u_0,v_0)$ ({\bf n}), though the two vectors are generally distinct.
Now consider the Darboux
frame $({\bf T},{\bf N},{\bf B})$ with ${\bf T}={\bf t},{\bf N}=\hat{\bf n}$
(the unit normal to $\Sigma$ in ${\bf R}(u_0,v_0)$), and
${\bf B}={\bf T}\wedge{\bf N}$.
Clearly, by a convenient rotation around ${\bf T}={\bf t}$, {\bf n} and
{\bf b} are carried to {\bf N} and {\bf B}, respectively. Calling
$\alpha(s)$ the rotation angle,
\be
\left(
\begin{array}{c}
{\bf T} \\
{\bf N} \\
{\bf B}
\end{array}
\right)=\left(
\begin{array}{ccc}
1 & 0 & 0 \\
0 & \cos\alpha & \sin\alpha \\
0 & -\sin\alpha & \cos\alpha
\end{array}
\right)\left(
\begin{array}{c}
{\bf t} \\
{\bf n} \\
{\bf b}
\end{array}
\right)\,.
\label{b-1}
\ee
Using the Frenet-Serret formulae, namely
\ba
\frac{{\rm d}{\bf t}}{{\rm d}s} &=& \kappa{\bf n}\,;
\nonumber \\
\frac{{\rm d}{\bf n}}{{\rm d}s} &=& -\kappa{\bf t}+\tau{\bf b}\,;
\nonumber \\
\frac{{\rm d}{\bf b}}{{\rm d}s} &=& -\tau{\bf n}\,,
\label{b-2}
\ea
where $\kappa$ is the curvature and $\tau$ is the torsion of $\ell$,
we easily get
\be
\left(
\begin{array}{c}
{\rm d}{\bf T}/{\rm d}s \\
{\rm d}{\bf N}/{\rm d}s \\
{\rm d}{\bf B}/{\rm d}s
\end{array}
\right)=\left(
\begin{array}{ccc}
0 & \kappa_n & \kappa_g \\
-\kappa_n & 0 & \tau_g \\
-\kappa_g & -\tau_g & 0
\end{array}
\right)\left(
\begin{array}{c}
{\bf T} \\
{\bf N} \\
{\bf B}
\end{array}
\right)\,,
\label{b-3}
\ee
where $\kappa_n=\kappa\cos\alpha$ is the normal curvature,
$\kappa_g=-\kappa\sin\alpha$ the geodetic curvature, and
$\tau_g=\tau+{\rm d}\alpha/{\rm d}s$ the geodetic torsion.

For the $u$-lines, if we identify {\bf T} with ${\bf R}_u$ then
${\bf B}={\bf T}\wedge{\bf N}=-{\bf R}_v$.
Similarly, for the $v$-lines, if we identify {\bf T} with ${\bf R}_v$
then ${\bf B}={\bf T}\wedge{\bf N}={\bf R}_u$.
We thus obtain:
\be
\frac{\partial\hat{\bf n}}{\partial u}=-\kappa_n^{(1)}{\bf R}_u-
\tau_g^{(1)}{\bf R}_v\,\,\,\,\,\,{\rm and}\,\,\,\,\,\,
\frac{\partial\hat{\bf n}}{\partial v}=\tau_g^{(2)}{\bf R}_u-
\kappa_n^{(2)}{\bf R}_v\,.
\label{b-4}
\ee
Moreover,
\ba
\frac{\partial{\bf R}_u}{\partial u} &=& -\kappa_g^{(1)}{\bf R}_v+
\kappa_n^{(1)}\hat{\bf n}\,\,\,\,\,\,{\rm and}\,\,\,\,\,\,
\frac{\partial{\bf R}_u}{\partial v}=-\kappa_g^{(2)}{\bf R}_v-
\tau_g^{(2)}\hat{\bf n}\,;
\nonumber \\
\frac{\partial{\bf R}_v}{\partial u} &=& \kappa_g^{(1)}{\bf R}_u+
\tau_g^{(1)}\hat{\bf n}\,\,\,\,\,\,{\rm and}\,\,\,\,\,\,
\frac{\partial{\bf R}_v}{\partial v}=\kappa_g^{(2)}{\bf R}_u+
\kappa_n^{(2)}\hat{\bf n}\,.
\label{b-5}
\ea
From ${\bf R}_{uv}={\bf R}_{vu}$, we derive
\be
\kappa_g^{(1)}{\bf R}_u+\kappa_g^{(2)}{\bf R}_v+\left(\tau_g^{(1)}+
\tau_g^{(2)}\right)\hat{\bf n}=0\,. 
\label{b-6}
\ee
Since ${\bf R}_u,{\bf R}_v$, and $\hat{\bf n}$ are linearly independent,
it necessarily follows that
\be
\kappa_g^{(1)}=\kappa_g^{(2)}=0\,\,\,\,\,\,{\rm and}\,\,\,\,\,\,
\tau_g^{(2)}=-\tau_g^{(1)}\equiv -\tau_g\,.
\label{b-7}
\ee
Since the geodetic curvature vanishes, our assumption that ${\bf R}_u$ and
${\bf R}_v$ are orthonormal vectors implies that the coordinate lines are
surface geodesics.

Finally, putting Eqs.\,(\ref{b-4}) and (\ref{b-7}) together
we promptly get Eq.\,(\ref{3-4}).

\section{Small fluctuations about a spherical interface}
\setcounter{equation}{0}
\renewcommand{\theequation}{C.\arabic{equation}}

We here provide the detailed derivation of Eq.\,(\ref{4-4})
for the free energy of a quasispherical interface $\Sigma$. The
starting point is the expansion of the relative amount of asphericity,
$\epsilon(\theta,\phi)$, in real spherical harmonics, Eq.\,(\ref{4-3}).
In view of the smallness of the expansion coefficients $x_{l,m}$,
the enclosed volume and area of $\Sigma$ can be approximated as
\be
{\cal V}[\Sigma]=\frac{1}{3}\int_\Sigma{\rm d}S\,{\bf r}\cdot\hat{\bf n}=
\frac{4}{3}\pi R_0^3+R_0^3\sum_{l>0,m}x_{l,m}^2\equiv\frac{4}{3}\pi
R_0^3\,f(\{x\})
\label{c-1}
\ee
and
\be
{\cal A}[\Sigma]=\int_\Sigma{\rm d}S=4\pi R_0^2
+\frac{R_0^2}{2}\sum_{l>0,m}\left(l^2+l+2\right)x_{l,m}^2\equiv
4\pi R_0^2\,g(\{x\})\,,
\label{c-2}
\ee
$f(\{x\})$ and $g(\{x\})$ being close-to-1 factors.
In writing the two formulae above we supposed $x_{0,0}=0$, which can
always be assumed by suitably redefining in $R(\theta,\phi)$
the radius $R_0$ and the other coefficients $x_{l,m}$.
In order to evaluate the mean curvature $H$, we start from
\be
\nabla\cdot\hat{\bf n}=\frac{1}{r^2}\frac{\partial(r^2\hat{n}_r)}{\partial r}+
\frac{1}{r\sin\theta}\frac{\partial}{\partial\theta}
(\sin\theta\,\hat{n}_\theta)+
\frac{1}{r\sin\theta}\frac{\partial\hat{n}_\phi}{\partial\phi}\,,
\label{c-3}
\ee
where
\ba
\hat{n}_r &=& 1-\frac{1}{2}\epsilon_\theta^2-
\frac{1}{2}\frac{\epsilon_\phi^2}{\sin^2\theta}\,;
\nonumber \\
\hat{n}_\theta &=& -\epsilon_\theta(1-\epsilon)\,;
\nonumber \\
\hat{n}_\phi &=& -\frac{\epsilon_\phi(1-\epsilon)}{\sin\theta}\,.
\label{c-4}
\ea
From that we get
\be
\nabla\cdot\widehat{\bf n}=\frac{2}{R(\theta,\phi)}
\left(1+\frac{1}{2}L^2\epsilon(\theta,\phi)-
\frac{1}{2}\epsilon(\theta,\phi)L^2\epsilon(\theta,\phi)\right)\,,
\label{c-5}
\ee
where
\be
L^2=-\frac{1}{\sin\theta}\frac{\partial}{\partial\theta}\left(\sin\theta
\frac{\partial}{\partial\theta}\right)-
\frac{1}{\sin^2\theta}\frac{\partial^2}{\partial\phi^2}\,.
\label{c-6}
\ee
Eventually, we obtain:
\ba
&& \int_\Sigma{\rm d}S\left(\sigma_m-2\sigma_m\delta_m H+2\lambda H^2\right)=
4\pi\sigma_m R_0^2+\frac{\sigma_m R_0^2}{2}\sum_{l>0,m}(l^2+l+2)x_{l,m}^2
\nonumber \\
&& -8\pi\sigma_m\delta_m R_0-\sigma_m\delta_m R_0\sum_{l>0,m}l(l+1)x_{l,m}^2+
8\pi\lambda+\frac{\lambda}{2}\sum_{l>1,m}l(l+1)(l-1)(l+2)x_{l,m}^2\,.
\nonumber \\
\label{c-7}
\ea
Finally, we specify the integral measure in (\ref{4-2}):
\be
\int{\cal D}\Sigma=\int_{-\infty}^{+\infty}\prod_{l>0,m}\left(
\frac{S}{s}\,{\rm d}x_{l,m}\right)\int_0^{+\infty}
\frac{{\rm d}R_0}{a}\,,
\label{c-8}
\ee
where $S=(36\pi)^{1/3}V^{2/3}$ is the area of the spherical surface of
volume $V$ and $s=4\pi a^2$. Equation (\ref{c-8}) follows from requiring
that the present theory (in fact the theory with an upper cutoff on $l$,
see below) should coincide with the continuum limit of the field theory for a
solid-on-solid (SOS) model with real heights defined on nodes uniformly
placed over a sphere of radius $\sqrt{S/(4\pi)}$.

To prove this, first observe that the equation for the generic $\Sigma$
entering in the functional integral is
$R-R_0=\sum_{l>0,m}R_0Y_{l,m}(\theta,\phi)x_{l,m}$.
Since $R_0=\sqrt{S/(4\pi)}$ up to terms ${\cal O}(x_{l,m}^2)$, the height
profile which the equation for $\Sigma$ corresponds to is
\be
h_i=\sum_{l>0,m}\sqrt{\frac{S}{4\pi}}Y_{l,m}(\Omega_i)x_{l,m}\,,
\label{c-9}
\ee
for $i=1,2,\ldots,n$ and $n=(l_{\rm max}+1)^2-1\simeq S/a^2$
(the necessity of an upper cutoff $l_{\rm max}$ on $l$ given by
the following Eq.\,(\ref{c-17}) will be motivated later).
The relation between the two theories passes through the identification
\be
\int\prod_{i=1}^n\frac{{\rm d}h_i}{a}\,\,\longleftrightarrow\,\,
\int\frac{J}{a^n}\prod_{l>0,m}{\rm d}x_{l,m}\,,
\label{c-10}
\ee
where $J$ is the Jacobian of the transformation (\ref{c-9}):
\be
J\equiv\det\left(\frac{\partial h_i}{\partial x_{l,m}}\right)=
\left(\frac{S}{4\pi}\right)^{n/2}\left|
\begin{array}{ccc}
Y_{1,-1}(\Omega_1) & \ldots & Y_{l_{\rm max},l_{\rm max}}(\Omega_1) \\
\vdots & \ddots & \vdots \\
Y_{1,-1}(\Omega_n) & \ldots & Y_{l_{\rm max},l_{\rm max}}(\Omega_n)
\end{array}
\right|\,.
\label{c-11}
\ee
Called $\Delta\Omega=4\pi/n$ the element of solid angle assigned to each node
$\Omega_i$, we have
\be
\sum_iY_{l,m}(\Omega_i)Y_{l',m'}(\Omega_i)\approx\frac{1}{\Delta\Omega}
\int{\rm d}^2\Omega\,Y_{l,m}(\Omega)Y_{l',m'}(\Omega)=\frac{1}{\Delta\Omega}
\delta_{l,l'}\delta_{m,m'}\,.
\label{c-12}
\ee
Hence, for sufficiently large $n$ the columns of the matrix (\ref{c-11}) are
mutually orthogonal
$n$-vectors. In order that every column vector be normalized, it suffices
to multiply the whole matrix in (\ref{c-11}) by $\sqrt{\Delta\Omega}$, thus
getting an orthogonal matrix (of unit determinant). Therefore, we find
\be
\frac{J}{a^n}=\left(\frac{S}{4\pi a^2}\right)^n\,,
\label{c-13}
\ee
which amounts to take $s=4\pi a^2$ in Eq.\,(\ref{c-8}). This completes
our proof.

We can now go on to compute the partition function (\ref{4-2}).
We first calculate
the integral on $R_0$ by rearranging the delta function in $Z_s$ as
\be
\delta\left(\frac{4}{3}\pi R_0^3\,f(\{x\})-V\right)=
\frac{\delta\left(R_0-\left[4\pi f(\{x\})/(3V)\right]^{-1/3}\right)}
{(36\pi)^{1/3}V^{2/3}f(\{x\})^{1/3}}\,.
\label{c-14}
\ee
After doing the trivial integral on $R_0$, we remain with a factor
$f(\{x\})^{-1/3}$ which, within a quadratic theory, can be treated as follows:
\be
f(\{x\})^{-1/3}=\left(1+\frac{3}{4\pi}\sum_{l>0,m}x_{l,m}^2\right)^{-1/3}
\simeq 1-\frac{1}{4\pi}\sum_{l>0,m}x_{l,m}^2\simeq
\exp\left\{-\frac{1}{4\pi}\sum_{l>0,m}x_{l,m}^2\right\}\,.
\label{c-15}
\ee
In the end, we arrive at a Gaussian integral which is readily computed:
\ba
Z_s &=& (36\pi)^{-1/3}\left(\frac{V}{a^3}\right)^{-2/3}
\exp\left\{\beta\rho_s|\Delta\mu|V-\beta\sigma_m S-8\pi\beta\lambda+
8\pi\beta\sigma_m\delta_m\left(\frac{3V}{4\pi}\right)^{1/3}\right\}
\nonumber \\
&\times& \left(\frac{2\pi S}{s}\right)^3
\prod_{l>1}\left\{\left(\frac{s}{2\pi S}\right)^2
\left[1+\frac{\beta\sigma_m S}{2}(l^2+l-2)+
2\pi\beta\lambda\,l(l+1)(l-1)(l+2)\right.\right.
\nonumber \\
&-& \left.\left.4\pi\beta\sigma_m\delta_m
\sqrt{\frac{S}{4\pi}}(l^2+l-2)\right]\right\}^{-(l+1/2)}\,.
\label{c-16}
\ea
Without a proper ultraviolet cutoff $l_{\rm max}$ the $l$ sum in
$\ln Z_s$ does not converge. This is a typical occurrence for
field theories on the continuum, which do not consider the
granularity of matter at the most fundamental level.
We fix $l_{\rm max}$ by requiring that the total number of $(l,m)$
modes be equal to the average number of SOS heights/atoms on the cluster
surface. It thus follows:
\be
l_{\rm max}=\frac{\sqrt{S}}{a}-1\,.
\label{c-17}
\ee
With this cutoff, the surface free energy becomes $F_s=\sigma(S)S$,
with an interface tension $\sigma(S)$ dressed by thermal fluctuations:
\ba
\sigma(S) &=& \sigma_m+\frac{k_BT}{2S}\sum_{l=2}^{\sqrt{S}/a-1}(2l+1)
\ln\left[A+B(l^2+l-2)+C(l^2+l-2)^2\right]
\nonumber \\
&-& 2\sigma_m\delta_m\left(\frac{4\pi}{S}\right)^{1/2}
-2\frac{k_BT}{S}\ln\left(\frac{S}{a^2}\right)
-3\frac{k_BT}{S}\ln\left(\frac{2\pi a^2}{s}\right)+
\frac{8\pi\lambda}{S}\,.
\label{c-18}
\ea
The quantities $A,B$, and $C$ in Eq.\,(\ref{c-18}) are given by
\be
A=\frac{A_0}{S^2}\,,\,\,\,B=\frac{2C_0}{S^2}+\frac{D_0}{S\sqrt{S}}+
\frac{B_0}{S}\,,\,\,\,C=\frac{C_0}{S^2}\,,
\label{c-19}
\ee
where
\be
A_0=\frac{s^2}{4\pi^2}\,,\,\,\,B_0=\frac{\beta\sigma_m s^2}{8\pi^2}\,,
\,\,\,C_0=\frac{\beta\lambda s^2}{2\pi}\,,\,\,\,
D_0=-\frac{\beta\sigma_m\delta_m s^2}{2\pi\sqrt{\pi}}\,.
\label{c-20}
\ee
By the Euler-Mac Laurin formula, the residual sum in Eq.\,(\ref{c-18})
can be evaluated explicitly. After a tedious and rather lengthy derivation,
we obtain (for $\lambda\neq 0$):
\ba
&& \sigma(S)=\sigma_m+\frac{k_BT}{2a^2}\left[\ln\frac{B_0}{a^2e^2}+
\left(1+\frac{B_0a^2}{C_0}\right)\ln\left(1+\frac{C_0}{B_0a^2}\right)\right]
\nonumber \\
&& +\left[-2\sigma_m\delta_m+\frac{k_BTD_0}{4C_0\sqrt{\pi}}
\ln\left(1+\frac{C_0}{B_0a^2}\right)\right]\left(\frac{4\pi}{S}\right)^{1/2}-
\frac{7}{6}k_BT\frac{\ln(S/a^2)}{S}
\nonumber \\
&& +\left[8\pi\beta\lambda-3\ln\frac{2\pi a^2}{s}-
\frac{11}{6}\ln\frac{B_0}{a^2}+3-\frac{5}{3}\ln 2-\frac{25}{96}+
\frac{121}{46080}\right.
\nonumber \\
&& +\frac{D_0a}{4C_0}-\frac{D_0^2}{4B_0C_0}-
\frac{1}{6}\ln\left(\frac{B_0}{a^2}+\frac{C_0}{a^4}\right)
+\frac{1}{8C_0(B_0a^2+C_0)^2}
\nonumber \\
&& \times\left(
-4B_0C_0D_0a^3-18B_0C_0^2a^2-2C_0^2D_0a-\frac{28}{3}C_0^3-
\frac{26}{3}B_0^2C_0a^4\right.
\nonumber \\
&& \left.\left.-2B_0^2D_0a^5+2B_0D_0^2a^4+2C_0D_0^2a^2
\right)\right]\frac{k_BT}{S}\,,
\label{c-21}
\ea
up to terms $o(S^{-1})$.
We wrote a computer code to evaluate the sum in (\ref{c-18}) numerically
for large $S$, and so checked that every single term in Eq.\,(\ref{c-21})
is indeed correct.

\section{Anisotropic-interface models of nucleation}
\setcounter{equation}{0}
\renewcommand{\theequation}{D.\arabic{equation}}

We here show that nonperturbative corrections to CNT do also arise when
the interface tension is infinitely anisotropic. In this case, the
admissible cluster shapes are all regular and the functional integral
(\ref{4-2}) is greatly simplified, reducing to a standard integral over
the few independent variables which concur to define the allowed clusters.
The terms in (\ref{3-15}) beyond the surface-tension term do also
contribute to the total surface free energy if anisotropy is strong
but not infinitely so.

Our argument goes as follows.
Let us, for instance, consider the interface tension (\ref{4-5}).
For $M\gg 1$, we expect that the leading contribution to the functional
integral (\ref{4-2}) be given by rectangular cuboids with slightly
rounded edges and vertices. Since we are only interested in making
a rough estimation of the relative magnitude of each contribution to $H_s$,
we assume that the surface of a rounded edge is one fourth of a cylindrical
surface ($H=1/a$) whereas that of a rounded vertex is an octant of a sphere
($H=2/a$), $a$ being a microscopic diameter. Also observe that:
$\sigma_{100}\sim k_BT_m/a^2$; the average value of $\sigma(\hat{\bf n})$
on an edge or vertex is $\sim M\sigma_{100}$; the Tolman length is
$\delta_m\sim a$; $\lambda$ is roughly $\kappa/c$ times $\sigma_m$,
hence $\lambda\sim\sigma_m a^2$.
We now decompose (\ref{3-15}) into the sum of three integrals,
respectively over faces, edges, and vertices.
Denoting $l_1,l_2,l_3$ (all much larger than $a$) the side lengths of
the cuboid if its edges and vertices were taken to be sharp, the integral
over faces is practically equal to $2\sigma_{100}(l_1l_2+l_1l_3+l_2l_3)\sim
k_BT_m(l_1l_2+l_1l_3+l_2l_3)/a^2$;
up to a factor of order one, the integral over edges is given by
$Mk_BT_m(l_1+l_2+l_3)/a$;
finally, the integral over vertices is of the order of $Mk_BT_m$.
We then see that, when $M\rightarrow\infty$ for fixed $a$,
only faces contribute to the integral (\ref{4-2}), while edges
and vertices would only matter if $M$ were finite.

In general terms, from the knowledge of the ``Wulff plot''
$\sigma(\hat{\bf n})$, the equilibrium cluster shape follows from the
so-called Wulff construction~\cite{Wulff}:
(i) draw the planes perpendicular to the unit vectors $\hat{\bf n}$ and at
a distance $\sigma(\hat{\bf n})$ away from the origin; (ii) for each plane,
discard the half-space of $\mathbb{R}^3$ that lies on the far side of the
plane from the origin. The convex region consisting of the intersection
of the retained half-spaces is the cluster of lowest surface energy.
When $\sigma(\hat{\bf n})$ is smooth, this Wulff cluster is bounded by
part of the envelope of the planes; the parts of the envelope not bounding
the convex body -- the ``ears'' or ``swallowtails'' which are e.g. visible
in Figs.\,3 and 6 below -- are unphysical.

\subsection{Rectangles}

In two dimensions, a rudimentary model of nucleation is that which only
allows for rectangular clusters. This is relevant for
two-dimensional crystals of square symmetry, and could be obtained
from a smooth interface tension of the form
\be
\sigma(\phi)=\sigma_{10}\left[1+M\sin^2(2\phi)\right]
\label{d-1}
\ee
upon taking the infinite-$M$ limit. In Eq.\,(\ref{d-1}), $\phi$ is the
polar angle of the normal vector while $\sigma_{10}$ is the free energy
of the cheapest, $(10)$ facet. As $M\rightarrow\infty$,
all normal directions different from [10], [01], [$\overline{1}$0], and
[0$\overline{1}$] are excluded from the equilibrium cluster shape and a
perfectly square surface is obtained. This is illustrated in Fig.\,3 for
three values of $M$; here and elsewhere, the envelope of perpendicular
planes is given, in parametric terms, by the equations~\cite{Burton}:
\ba
x&=&\cos\phi\,\sigma(\phi)-\sin\phi\,\sigma'(\phi)\,;
\nonumber \\
y&=&\sin\phi\,\sigma(\phi)+\cos\phi\,\sigma'(\phi)\,.
\label{d-2}
\ea

%
%   FIGURE 3
%
\begin{figure}
\includegraphics[width=10cm]{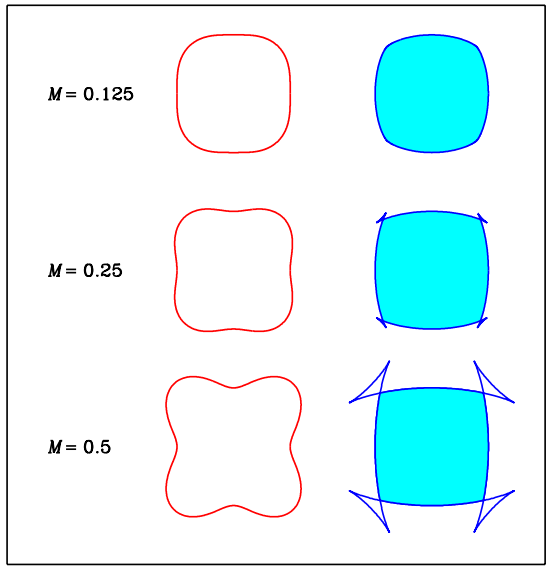}
\caption{(Color online). Two-dimensional Wulff construction for a square
cluster. Left: Polar plot of the interface tension $\sigma$ in
Eq.\,(\ref{d-1}) (red curves) as a function of the normal
$\hat{\bf n}=\cos\phi\,\hat{\bf x}+\sin\phi\,\hat{\bf y}$ to a cluster face,
for three distinct values of $M$. Right: Equilibrium cluster shape as the
envelope of the family of perpendicular planes (blue curves). In this case,
the cluster of minimum surface energy (colored in cyan) has curved faces,
but sharp corners. The envelope continues beyond the corners, but these
parts have no physical meaning.}
\label{fig3}
\end{figure}

Note that the edge fluctuations deforming the square in a rectangle are still
allowed by (\ref{d-1}) in the infinite-$M$ limit, since rectangles and cubes
share the same type of facets. Hence, assuming that only rectangular shapes
have a non-zero Boltzmann weight in the functional integral (\ref{4-2}),
the surface free energy reduces to:
\be
F_s(V)=-\frac{1}{\beta}\ln\int_0^{+\infty}{\rm d}a
\int_0^{+\infty}{\rm d}b\,e^{-2\beta\sigma(a+b)}\delta(ab-V)
\label{d-3}
\ee
We easily find:
\ba
\beta\Delta G(V) &\equiv& -\beta gV+\beta F_s(V)
\nonumber \\
&=& -\beta gV+4\beta\sigma\sqrt{V}-
\ln\int_{-\infty}^{+\infty}{\rm d}x\,\exp\left\{-\tau(\cosh x-1)\right\}\,,
\nonumber \\
&=& -\beta gV+4\beta\sigma\sqrt{V}-\ln\left\{2e^\tau\int_1^{+\infty}{\rm d}t\,
\frac{e^{-\tau t}}{\sqrt{t^2-1}}\right\}
\nonumber \\
&=& -\beta gV+4\beta\sigma\sqrt{V}-\ln\left\{2e^\tau K_0(\tau)\right\}\,,
\label{d-4}
\ea
where $g$ is proportional to the supersaturation, $\tau=4\beta\sigma\sqrt{V}$,
and $K_0$ is a modified Bessel function of the second kind. The last term in
Eq.\,(\ref{d-4}) is the full correction to CNT as formulated for squares.
A typical profile of $\beta\Delta G(V)$ is plotted in Fig.\,4.

%
%    FIGURE 4
%
\begin{figure}
\includegraphics[width=10cm]{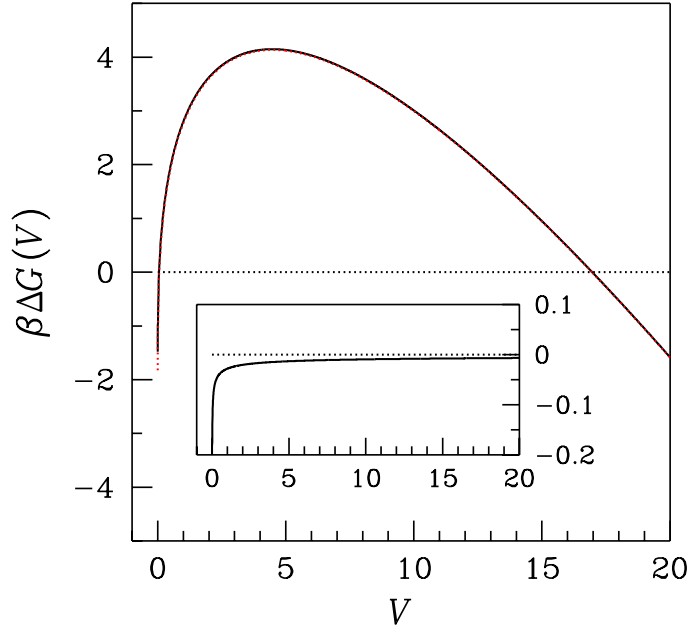}
\caption{(Color online). Rectangles: $\beta\Delta G(V)$ vs. $V$
for $\beta g=\beta\sigma=1$ (black solid line). The dotted red line
corresponds to the approximant (\ref{d-8}). Inset, the difference between
(\ref{d-8}) and $\beta\Delta G(V)$.}
\label{fig4}
\end{figure}

At variance with CNT, $\Delta G(V)$ shows a weak divergence to $-\infty$ for
$V\rightarrow 0$, due to the absence of a lower cutoff volume.
For $\tau\ll 1$,
\be
K_0(\tau)=-\ln(\tau/2)-\gamma+{\cal O}(\tau^2\ln\tau)\,,
\label{d-5}
\ee
with $\gamma=0.5772\ldots$ (Eulero-Mascheroni constant). Hence, the singular
behavior of $\beta\Delta G(V)$ for small $V$ is of the kind
\be
\beta\Delta G(V)\simeq
-\ln\left\{-\ln\left(2\beta\sigma\sqrt{V}\right)\right\}\,.
\label{d-6}
\ee
Conversely, for large $\tau$ values,
\be
K_0(\tau)\sim\sqrt{\frac{\pi}{2\tau}}\,e^{-\tau}\,,
\label{d-7}
\ee
and we obtain
\be
\beta\Delta G(V)\simeq -\beta gV+4\beta\sigma\sqrt{V}+
\frac{1}{2}\ln\left(4\beta\sigma\sqrt{V}\right)-\frac{1}{2}\ln(2\pi)\,.
\label{d-8}
\ee
The goodness of the approximation (\ref{d-8}) can be judged from the
inset of Fig.\,4, which shows that the approximation is accurate for
all values of $V$ but for the smallest ones.

%
%    FIGURE 5
%
\begin{figure}
\centering
\includegraphics[width=10cm]{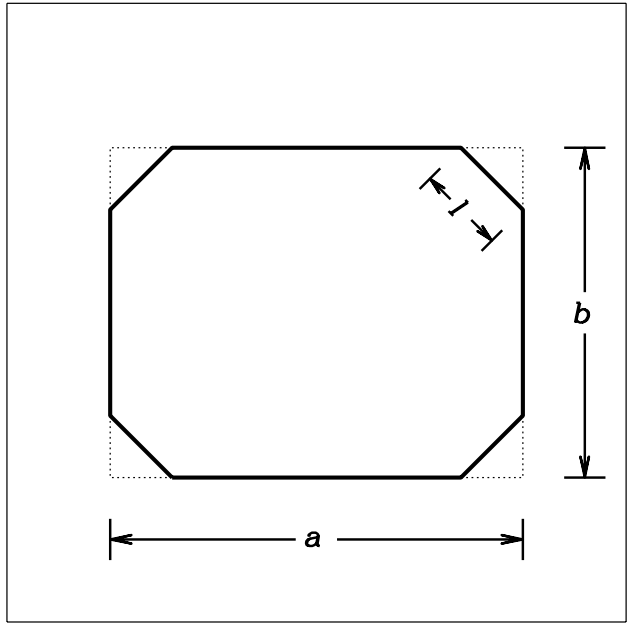}
\caption{A truncated rectangle. For fixed $a$ and $b$, the maximum $\ell$
value is $\ell_{\rm max}=(1/\sqrt{2})\min\{a,b\}$.}
\label{fig5}
\end{figure}

\subsection{Truncated rectangles}

In order to study the effects on nucleation of a more complicate type of
interface-tension anisotropy,
we further enrich our book of patterns, passing from rectangles to truncated
rectangles. By the name of truncated rectangle we mean the octagon
represented in Fig.\,5. This occurs when the cost of (11) and
equivalent facets is of the same order of $\sigma_{10}$, while all other
facets are much higher in energy and can be ruled out.

%
%   FIGURE 6
%
\begin{figure}
\includegraphics[width=10cm]{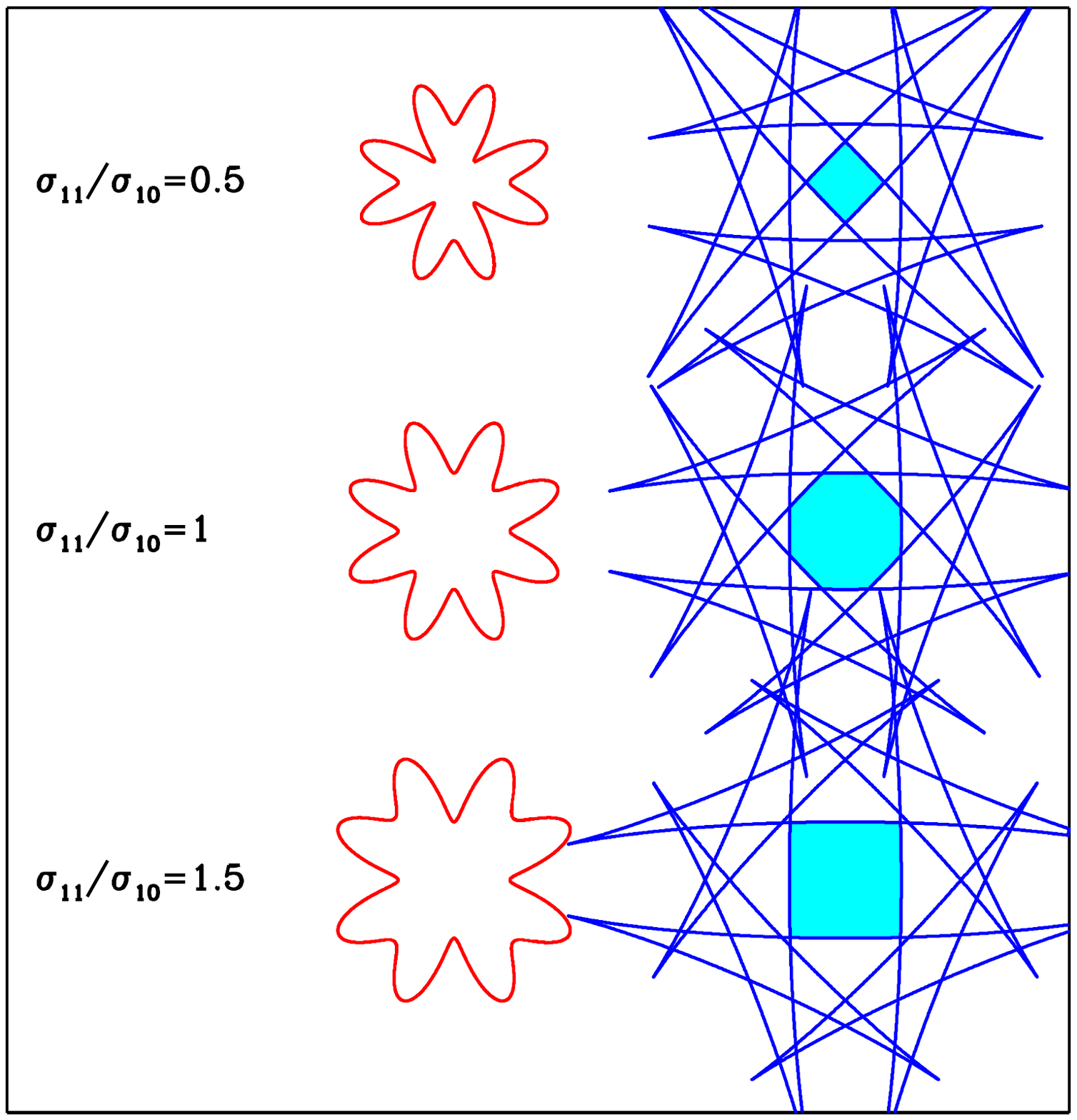}
\caption{(Color online). Two-dimensional Wulff construction for the $\sigma$
model at Eq.\,(\ref{d-9}), with $M=1$. Left: Wulff plot (red) for three values
of $\sigma_{11}/\sigma_{10}$. Right: Equilibrium cluster shape (the
boundary of the cyan-colored region). The parts of the envelope of the family
of perpendicular planes beyond the corners of the cluster are unphysical.}
\label{fig6}
\end{figure}

A Wulff plot giving origin to truncated squares is:
\be
\sigma(\phi)=\sigma_{10}\left[1+\left(\frac{\sigma_{11}}{\sigma_{10}}-1\right)
\sin^2(2\phi)+M\sin^2(4\phi)\right]
\label{d-9}
\ee
with infinite $M$ (see Fig.\,6). The polar plot of (\ref{d-9}) for finite
$M$ is a smoothed
eight-pointed star with hollows at the normal directions satisfying
$\sin(4\phi)=0$. Depending on the ratio of $\sigma_{11}$ to $\sigma_{10}$,
the equilibrium cluster shape shows (i) just (11) facets
($\sigma_{11}/\sigma_{10}\le\sqrt{2}/2$); (ii) both (11) and (10) facets
($\sqrt{2}/2<\sigma_{11}/\sigma_{10}<\sqrt{2}$); (iii) just (10) facets
($\sigma_{11}/\sigma_{10}\ge\sqrt{2}$).

In order to prove this, we observe that, for fixed $a,b$, and $\ell$
(with $\ell\le\ell_{\rm max}\equiv(1/\sqrt{2})\min\{a,b\}$), the ``volume''
and ``area'' of the truncated rectangle are given respectively by
$V=ab-\ell^2$ and $A=A_{11}+A_{10}$,
with $A_{11}=4\ell$ and $A_{10}=2(a+b-2\sqrt{2}\ell)$,
leading to a surface energy of
\be
E_s=4\sigma_{11}\ell+2\sigma_{10}(a+b-2\sqrt{2}\ell)\,.
\label{d-10}
\ee
To determine the cluster shape at zero temperature, $E_s$ should be minimized
as a function of $a,b$, and $\ell$ under the constraint of a fixed
$ab-\ell^2\,(=V)$.
Setting $a=x\sqrt{V}$ and $b=y\sqrt{V}$ (with $x,y>0$), we are led to minimize
$4(\sigma_{11}-\sqrt{2}\sigma_{10})\sqrt{xy-1}+2\sigma_{10}(x+y)$
as a function of $x$ and $y$. By a straightforward calculation we find:
\ba
&& x=y=1\,\,\,\,(\ell=0)\,,\,\,{\rm for}\,\,
\frac{\sigma_{11}}{\sigma_{10}}\ge\sqrt{2}\,;
\nonumber \\
&& x=y=\frac{1}{\sqrt{1-\left(\sqrt{2}-\sigma_{11}/\sigma_{10}\right)^2}}
\,\,\,\,\left(\frac{\ell}{\sqrt{V}}=\frac{\sqrt{2}-\sigma_{11}/\sigma_{10}}
{\sqrt{1-\left(\sqrt{2}-\sigma_{11}/\sigma_{10}\right)^2}}\right)
\,,\,\,{\rm for}\,\,
\frac{\sqrt{2}}{2}<\frac{\sigma_{11}}{\sigma_{10}}<\sqrt{2}\,;
\nonumber \\
&& x=y=\sqrt{2}\,\,\,\,\left(\frac{\ell}{\sqrt{V}}=1\right)\,,
\,\,{\rm for}\,\,\frac{\sigma_{11}}{\sigma_{10}}\le\frac{\sqrt{2}}{2}\,.
\label{d-11}
\ea
The equilibrium cluster shape is then a square (respectively, a 45-degree
tilted square) for $\sigma_{11}/\sigma_{10}$ values larger than $\sqrt{2}$
(smaller than $\sqrt{2}/2$), while being a truncated square otherwise
(see Fig.\,6).

Now going to the nucleation model for truncated rectangles, the surface free
energy reads:
\be
F_s(V)=-\frac{1}{\beta}\ln\iint_0^{+\infty}{\rm d}a\,{\rm d}b
\int_0^{\ell_{\rm max}}\frac{{\rm d}\ell}{\ell_0}\,
e^{-4\beta\sigma_{11}\ell}e^{-2\beta\sigma_{10}(a+b-2\sqrt{2}\ell)}
\,\delta(ab-\ell^2-V)\,,
\label{d-12}
\ee
where $\ell_0$ is an arbitrary length. By integrating the delta out,
we obtain:
\ba
\beta\Delta G(V) &=& -\beta gV+4\beta\sigma_{10}\sqrt{V}-
\frac{1}{2}\ln\left(\frac{V}{\ell_0^2}\right)
\nonumber \\
&-& \ln\int_{-\infty}^{+\infty}{\rm d}x\int_0^{+\infty}{\rm d}y\,
\Theta\left(\min\{e^x,(1+y^2)e^{-x}\}-\sqrt{2}y\right)
\nonumber \\
&\times& \exp\left\{-\left(\tau_{11}-\sqrt{2}\tau_{10}\right)y\right\}
\exp\left\{-\frac{\tau_{10}}{2}\left(e^x+(1+y^2)e^{-x}-2\right)\right\}\,,
\label{d-13}
\ea
where $\Theta$ is Heaviside's function, $\tau_{10}=4\beta\sigma_{10}\sqrt{V}$,
and $\tau_{11}=4\beta\sigma_{11}\sqrt{V}$. For $\beta\sigma_{10}=1,\ell_0=1$,
and $\sigma_{11}/\sigma_{10}=0.5,1,2,20,200$, the plot of (\ref{d-13})
is reported in Fig.\,7. Note that a precritical minimum shows up for
any finite value of $\sigma_{11}/\sigma_{10}$, which moves toward zero
upon increasing the interface-tension anisotropy. An even more complex
behavior is seen for $\sigma_{11}/\sigma_{10}=0.5$, where a bump emerges
beyond the critical maximum.

%
%    FIGURE 7
%
\begin{figure}
\centering
\includegraphics[width=10cm]{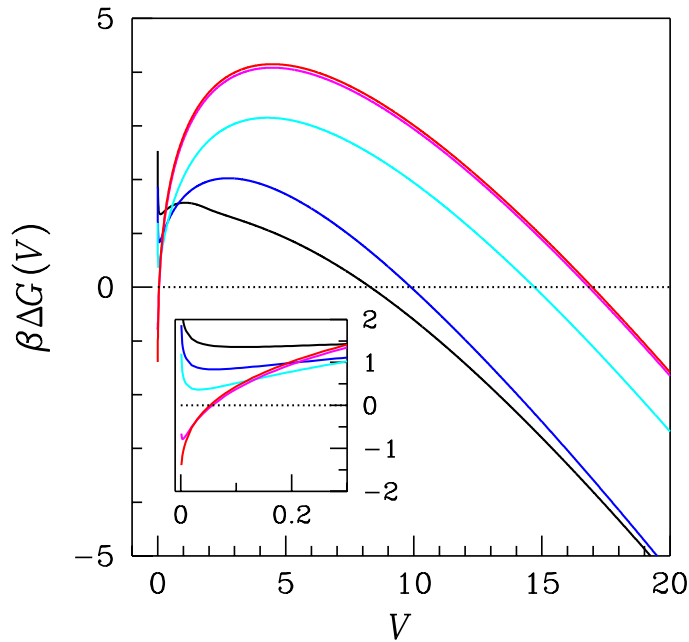}
\caption{(Color online). Truncated rectangles: $\beta\Delta G(V)$ vs. $V$
for $\beta g=\beta\sigma_{10}=1$ e $\ell_0=1$. A constant of
$\ln(4\beta\sigma_{11})$ was subtracted from $\beta\Delta G(V)$ in order
to garantee the confluence of its plot to that for rectangles, in the limit
$\sigma_{11}/\sigma_{10}\rightarrow +\infty$. A number of
$\sigma_{11}/\sigma_{10}$ values are considered: 0.5 (black), 1 (blue),
2 (cyan), 20 (magenta), and 200 (red, practically indistinguishable from
the rectangular case).
In the inset, we zoom on the small-$V$ region, evidencing the singular
behavior of $\Delta G(V)$ for $V\rightarrow 0$. Apparently, for all
finite $\sigma_{11}$ values, the curve blows up to $+\infty$
rather than to $-\infty$, as instead occurs for rectangles.}
\label{fig7}
\end{figure}

When $\sigma_{11}\gg\sigma_{10}$, it is natural to expect that the model
of truncated rectangles reduces to the rectangular-cluster model.
This can be proved analytically starting from Eq.\,(\ref{d-12}).
First, $a,b$, and $\ell$ are rescaled by dividing by $\sqrt{V}$; then
one observes that
\be
\sqrt{V}e^{-4\beta\sigma_{11}\sqrt{V}\ell}\approx\frac{1}{2\beta\sigma_{11}}
\delta(\ell)\,.
\label{d-14}
\ee
Hence, aside from a constant equal to $\ln(4\beta\sigma_{11})$, the
$\beta\Delta G(V)$ function for truncated rectangles merges,
for very large $\sigma_{11}/\sigma_{10}$, into the analogous function
for rectangles. This fact is shown numerically in the inset of Fig.\,7.

\subsection{Rectangular cuboids}

When the Wulff plot is as in Eq.\,(\ref{4-5}) with infinite $M$,
the only admissible shapes are rectangular cuboids.
Denoting $a,b$, and $c$ the edges of a cuboid, the $V$-dependent
surface free energy is defined as
\ba
\beta F_s(V) &=& -\ln\iiint_0^{+\infty}
{\rm d}a\,{\rm d}b\,{\rm d}c\,e^{-2\beta\sigma(ab+ac+bc)}\delta(abc-V)
\nonumber \\
&=& -\ln\iint_0^{+\infty}{\rm d}a\,{\rm d}b\,\frac{1}{ab}
\exp\left\{-\frac{\tau}{3}\left(ab+\frac{a+b}{ab}\right)\right\}\,,
\label{d-15}
\ea
where $\tau=6\beta\sigma V^{2/3}$. With another change of variables,
we arrive at
\be
\beta\Delta G(V)=-\beta gV+6\beta\sigma V^{2/3}-
\ln\iint_{-\infty}^{+\infty}{\rm d}x\,{\rm d}y\,
\exp\left\{-\frac{\tau}{3}\left(e^{x+y}+e^{-x}+e^{-y}-3\right)\right\}\,.
\label{d-16}
\ee
The above formula is well suited for the numerical evaluation of $\Delta G(V)$.
For $\beta g=\beta\sigma=1$, the profile of $\beta\Delta G(V)$ is plotted
in Fig.\,8.

%
%    FIGURE 8
%
\begin{figure}
\includegraphics[width=10cm]{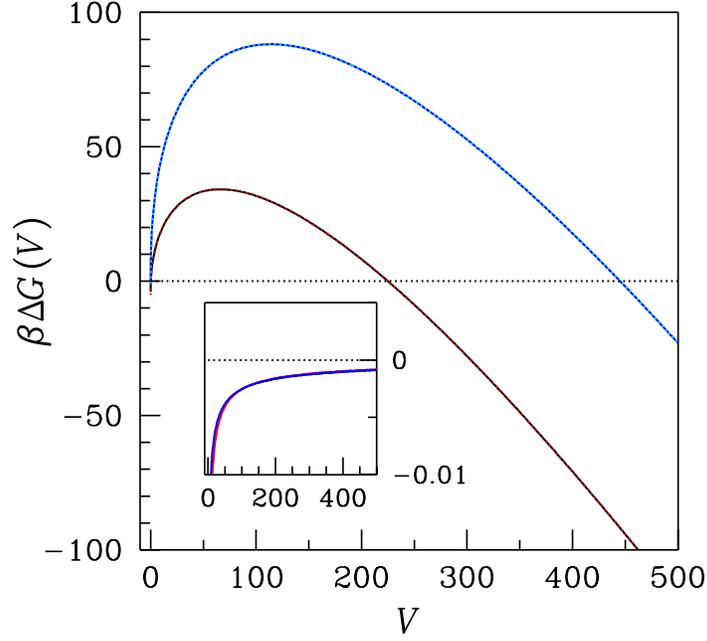}
\caption{(Color online). Rectangular cuboids: $\beta\Delta G(V)$ vs. $V$ for
$\beta g=\beta\sigma=1$ (black solid line). In blue, the same function
when we include the cost of the edges ($\beta g=\beta\sigma=\beta\nu=1$).
The red dotted line is the approximant (\ref{d-27}) while the cyan dotted
line is the approximant (\ref{d-31}). In the inset, we plot in red the
difference between the approximation (\ref{d-27}) and (\ref{d-16}), and
in blue the difference between (\ref{d-31}) and (\ref{d-28}).}
\label{fig8}
\end{figure}

In order to discover the analytic behavior of $\Delta G(V)$ at small
and at large $V$'s, we should further elaborate on Eq.\,(\ref{d-16}).
Setting $a+b=x$ and $ab=y$ in (\ref{d-15}), $a$ and $b$ are the solutions
to the equation $t^2-xt+y=0$, whose discriminant is non-negative for
$x\ge 2\sqrt{y}$. Moreover, the Jacobian of the transformation is
$1/\sqrt{x^2-4y}$. Hence, we get $\beta F_s(V)\equiv -\ln I(V)$ with
\be
I=2\int_0^{+\infty}{\rm d}y\int_{2\sqrt{y}}^{+\infty}{\rm d}x\,
\frac{\exp\{-(\tau/3)(y+x/y)\}}{y\sqrt{x^2-4y}}\,.
\label{d-17}
\ee
With the further transformations $x\rightarrow z=\sqrt{x^2-4y}$ and
$z\rightarrow w=z/y$, we eventually obtain:
\ba
I &=& 2\int_0^{+\infty}{\rm d}y\,\frac{e^{-(\tau/3)y}}{y}\int_0^{+\infty}
{\rm d}z\,\frac{\exp\{-(\tau/3)\sqrt{z^2+4y}/y\}}{\sqrt{z^2+4y}}
\nonumber \\
&=& 2\int_0^{+\infty}{\rm d}y\,\frac{e^{-(\tau/3)y}}{y}\int_0^{+\infty}
{\rm d}w\,\frac{\exp\left\{-\sqrt{w^2+\frac{4\tau^2}{9y}}\right\}}
{\sqrt{w^2+\frac{4\tau^2}{9y}}}\,.
\label{d-18}
\ea
Since
\be
\int_0^{+\infty}{\rm d}x\,\frac{\exp\{-\sqrt{x^2+c^2}\}}{\sqrt{x^2+c^2}}=
\int_c^{+\infty}{\rm d}t\,\frac{e^{-t}}{\sqrt{t^2-c^2}}=K_0(c)\,,
\label{d-19}
\ee
we finally find:
\be
I=2\int_0^{+\infty}{\rm d}y\,\frac{e^{-(\tau/3)y}}{y}
K_0\left(\frac{2\tau}{3\sqrt{y}}\right)=4\int_0^{+\infty}{\rm d}x\,
\exp\left\{-\frac{4\tau^3}{27x^2}\right\}\frac{K_0(x)}{x}\,.
\label{d-20}
\ee
In Eq.\,(\ref{d-20}) we recognize a particular Meijer function,
$G_{03}^{30}((\tau/3)^3|0,0,0)$, whose behavior at small $\tau$ is:
\be
\frac{9}{2}(\ln\tau)^2+9(\gamma-\ln 3)\ln\tau+{\cal O}(1)\,.
\label{d-21}
\ee
From the above, we can draw the main singular term in $\beta\Delta G(V)$ at
small $V$, that is
\be
\beta\Delta G(V)\simeq -2\ln\left(-\ln(6\beta\sigma V^{2/3})\right)\,,
\label{d-22}
\ee
which is similar to (\ref{d-6}).

The large-$V$ behavior of $\Delta G(V)$ can also be obtained from
Eq.\,(\ref{d-20}). For $\tau\gg 1$, we are allowed to replace $K_0(x)$
with Eq.\,(\ref{d-7}) and thus estimate $I$ through the integral
\be
I_\infty=2\sqrt{2\pi}\int_0^{+\infty}{\rm d}x\,
\frac{\exp\left\{-x-\frac{4\tau^3}{27x^2}\right\}}{x^{3/2}}\,.
\label{d-23}
\ee
Suspecting a dominant term of $\tau$ in $-\ln I_\infty$, we consider
\be
e^\tau I_\infty=2\sqrt{\frac{2\pi}{\tau}}\int_0^{+\infty}{\rm d}z\,
z^{-3/2}\exp\left\{\tau\left(1-z-\frac{4}{27z^2}\right)\right\}\,.
\label{d-24}
\ee
In order to compute the asymptotic behavior of (\ref{d-24}), we use
the Laplace method. The maximum of the concave function
$\phi(z)=1-z-(4/27)z^{-2}$ falls at $c=2/3$, with $\phi(c)=0$ and
$\phi''(c)=-9/2$. Since, for any $a<c<b$,
\be
\int_a^b{\rm d}z\,f(z)e^{\tau\phi(z)}\sim\frac{\sqrt{2\pi}f(c)e^{\tau\phi(c)}}
{\sqrt{-\tau\phi''(c)}}\,,
\label{d-25}
\ee
the asymptotic behavior of $I$ reads:
\be
I\sim 2\pi\sqrt{3}\,\frac{e^{-\tau}}{\tau}
\label{d-26}
\ee
and
\be
\beta\Delta G(V)\sim -\beta gV+6\beta\sigma V^{2/3}+
\ln(6\beta\sigma V^{2/3})-\ln(2\pi\sqrt{3})\,.
\label{d-27}
\ee
The last two terms in Eq.\,(\ref{d-27}) give the subleading corrections
to CNT as formulated for cubic clusters.
The quality of the approximation (\ref{d-27}) can be judged from the
inset of Fig.\,8, which shows a very good matching for all $V$'s except for
the smallest values, similarly to what occurs for rectangles (cf. Fig.\,4).

The calculation of $\Delta G$ can also be performed when a further energy
cost, $\nu$ per unit length, is assumed for the edges. Equation (\ref{d-16})
is then modified to
\ba
&& \beta\Delta G(V)=-\beta gV+6\beta\sigma V^{2/3}+12\beta\nu V^{1/3}
\nonumber \\
&-& \ln\iint_{-\infty}^{+\infty}{\rm d}x\,{\rm d}y\,
\exp\left\{-\frac{\tau_1}{3}\left(e^{x+y}+e^{-x}+e^{-y}-3\right)
-\frac{\tau_2}{3}\left(e^{-x-y}+e^x+e^y-3\right)\right\}\,,
\nonumber \\
\label{d-28}
\ea
where $\tau_1=6\beta\sigma V^{2/3}$ and $\tau_2=12\beta\nu V^{1/3}$.
By the same line of reasoning as followed above we arrive at
$\beta F_s\equiv -\ln I(V)$ with
\be
I=2\int_0^{+\infty}{\rm d}y\,\frac{e^{-(\tau_1/3)y-\tau_2/(3y)}}{y}
K_0\left(\frac{2\tau_1}{3\sqrt{y}}+\frac{2\tau_2}{3}y\right)\,.
\label{d-29}
\ee
Laplace method can still be invoked to extract the asymptotic behavior of $I$,
which turns out to be
\be
-\ln I\sim\tau_1+\tau_2+\ln(\tau_1+\tau_2)-\ln(2\pi\sqrt{3})\,.
\label{d-30}
\ee
From the above formula, we get
\be
\beta\Delta G(V)\sim-\beta gV+6\beta\sigma V^{2/3}+12\beta\nu V^{1/3}+
\ln(6\beta\sigma V^{2/3}+12\beta\nu V^{1/3})-\ln(2\pi\sqrt{3})\,.
\label{d-31}
\ee
In Fig.\,8, we compare the approximation (\ref{d-31}) with the exact value.
We see that the agreement is good for not too small $V$.

\subsection{\bf Ellipsoids}

Let us finally study the case of an ellipsoidal cluster.
Volume and area of an ellipsoid with semiaxes $a,b$, and $c$
are respectively given by
\ba
V &=& \frac{4}{3}\pi abc\,\,\,\,\,\,{\rm and}\,\,\,\,\,\,
A=2\pi\left(c^2+\frac{bc^2}{\sqrt{a^2-c^2}}F(\phi|m)+b\sqrt{a^2-c^2}
\,E(\phi|m)\right)\,,
\nonumber \\
&& (a\ge b>c\,\,{\rm and}\,\,a>b\ge c\,;\,\,A=4\pi c^2\,\,{\rm for}\,\,
a=b=c)
\label{d-32}
\ea
where
\be
m=\frac{a^2(b^2-c^2)}{b^2(a^2-c^2)}=\frac{1-c^2/b^2}{1-c^2/a^2}<1
\,\,\,\,\,\,{\rm and}\,\,\,\,\,\,\phi=\arcsin\frac{\sqrt{a^2-c^2}}{a}\,.
\label{d-33}
\ee
$F$ and $E$ are elliptic integrals of the first and second kind, respectively.
For $-\pi/2<\phi<\pi/2$, they are defined as
\be
F(\phi|m)\equiv\int_0^\phi{\rm d}x\,\frac{1}{\sqrt{1-m\sin^2x}}
\,\,\,\,\,\,{\rm and}\,\,\,\,\,\,
E(\phi|m)\equiv\int_0^\phi{\rm d}x\,\sqrt{1-m\sin^2x}\,.
\label{d-34}
\ee

%
%    FIGURE 9
%
\begin{figure}
\includegraphics[width=10cm]{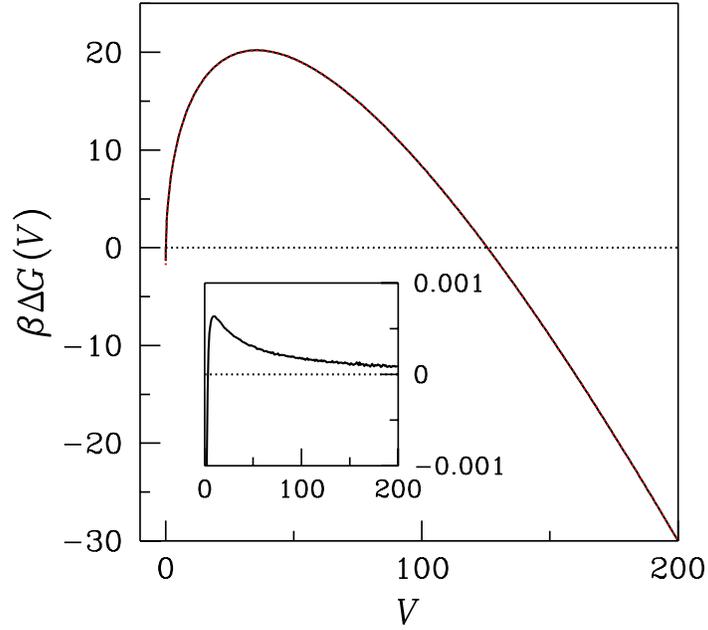}
\caption{(Color online). Ellipsoids: $\beta\Delta G(V)$ vs. $V$
for $\beta g=\beta\sigma=1$ (black solid line). The red dotted line is
the approximant (\ref{d-36}). Inset, the difference between
the large-$V$ estimate (\ref{d-36}) and the exact $\beta\Delta G(V)$.}
\label{fig9}
\end{figure}

Let now $A(a,b,c)$ be the surface area of an ellipsoid of semiaxes
$a,b$, and $c$ (not necessarily in descending order).
By the usual transformations, the surface free energy becomes
\ba
\beta F_s(V) &=& -\ln\iint_0^{+\infty}{\rm d}a\,{\rm d}b\,\frac{1}{ab}
\exp\left\{-\beta\sigma\left(\frac{3V}{4\pi}\right)^{2/3}
A\left(a,b,\frac{1}{ab}\right)\right\}+\ln\frac{4\pi}{3}
\nonumber \\
&=& \tau-\ln\iint_{-\infty}^{+\infty}{\rm d}q\,{\rm d}p\,
\exp\left\{-\frac{\tau}{4\pi}\left(A\left(e^q,e^p,e^{-q-p}\right)-4\pi\right)
\right\}+\ln\frac{4\pi}{3}\,,
\label{d-35}
\ea
where $\tau=\beta\sigma(36\pi)^{1/3}V^{2/3}$. To obtain $\beta\Delta G(V)$,
it is sufficient to add $-\beta gV$ to (\ref{d-35}).
For $\beta g=\beta\sigma=1$, the plot of this function is reported in Fig.\,9.
In the same figure, $\beta\Delta G(V)$ is compared with the asymptotic
estimate
\be
\beta\Delta G(V)\sim -\beta gV+\beta\sigma(36\pi)^{1/3}V^{2/3}+
\ln(\beta\sigma(36\pi)^{1/3}V^{2/3})-0.4849\,,
\label{d-36}
\ee
where the last two terms give the correction to CNT as formulated now
for spherical clusters.
Judging from the inset of Fig.\,9, which shows the difference between
the approximate and exact values of $\beta\Delta G(V)$, the estimate
(\ref{d-36}) is very good for all $V$'s except for the very small ones.

The strong similarity between (\ref{d-36}) and (\ref{d-27}),
together with the high accuracy with which they reproduce the profile
of $\beta\Delta G(V)$ for ellipsoids and cuboids respectively, indicates
that the difference between envisaging the nucleus as ellipsoidal rather
than cuboidal entirely lies in the value of $\sigma$, which for
an ellipsoid is $6/(36\pi)^{1/3}\simeq 1.241$ times the cuboidal one.
This occurs exactly as in CNT where the same relation holds between the
values of $\sigma$ for spheres and cubes.


\begin{thebibliography}{99}
\bibitem{Kelton} K. F. Kelton, {\em Solid State Physics}, vol.\,45, 75-90
(Academic, New York, 1991).

\bibitem{Kashchiev} D. Kashchiev, {\em Nucleation: Basic Theory with
Applications} (Butterworth-Heinemann, Oxford, 2000).

\bibitem{Vekilov} P. G. Vekilov, {\em Soft Matter} {\bf 6}, 5254 (2010).

\bibitem{Volmer} M. Volmer and A. Weber, {\em Z. Phys. Chem.}
{\bf 119}, 277 (1926).

\bibitem{Farkas} L. Farkas, {\em Z. Phys. Chem.} {\bf 125}, 239 (1927).

\bibitem{Becker} R. Becker and W. D\"{o}ring, {\em Ann. Phys.} (Leipzig)
{\bf 24}, 719 (1935).

\bibitem{Bagdassarian} See e.g. C. K. Bagdassarian and D. W. Oxtoby,
{\em J. Chem. Phys.} {\bf 100}, 2139 (1994).

\bibitem{Prestipino} S. Prestipino, A. Laio, and E. Tosatti,
{\em Phys. Rev. Lett.} {\bf }, (2012).

\bibitem{Dillmann} A. Dillmann and G. E. A. Meier, {\em J. Chem. Phys.}
{\bf 94}, 3872 (1991).

\bibitem{Shih} See e.g. W. H. Shih, Z. Q. Wang, X. C. Zeng, and D. Stroud,
{\em Phys. Rev. A} {\bf 35} 2611 (1987).

\bibitem{Cahn} J. W. Cahn and J. E. Hilliard,
{\em J. Chem. Phys.} {\bf 28}, 258 (1957).

\bibitem{Cahn2} J. W. Cahn and J. E. Hilliard,
{\em J. Chem. Phys.} {\bf 31}, 688 (1959).

\bibitem{Fisher} M. P. A. Fisher and M. Wortis, {\em Phys. Rev. B} {\bf 29},
6252 (1984).

\bibitem{Harrowell} P. Harrowell and D. W. Oxtoby, {\em J. Chem. Phys.}
{\bf 80}, 1639 (1984).

\bibitem{Shen} Y. C. Shen and D. W. Oxtoby, {\em J. Chem. Phys.} {\bf 105},
6517 (1996).

\bibitem{Kardar} See, for example, M. Kardar, {\em Statistical Physics of
Fields} (Cambridge University Press, 2007).

\bibitem{Block} B. J. Block, S. K. Das, M. Oettel, P. Virnau, and K. Binder,
{\em J. Chem. Phys.} {\bf 133}, 154702 (2010).

\bibitem{Tolman} R. C. Tolman, {\em J. Chem. Phys.} {\bf 17}, 333 (1949).

\bibitem{Filion} See e.g. L. Filion, M. Hermes, R. Ni, and M. Dijkstra,
{\em J. Chem. Phys.} {\bf 133}, 244115 (2010).

\bibitem{Zykova} T. Zykova-Timan, C. Valeriani, E. Sanz, D. Frenkel,
and E. Tosatti, {\em Phys. Rev. Lett.} {\bf 100}, 036103 (2008). 

\bibitem{Canham} P. Canham, {\em J. Theor. Biol.} {\bf 26}, 61 (1970).

\bibitem{Helfrich} W. Helfrich, {\em Z. Naturforsch. C} {\bf 28}, 693 (1973).

\bibitem{Kogon} H. S. Kogon and D. J. Wallace, {\em J. Phys. A} {\bf 14},
L527 (1981).

\bibitem{Kassner} K. Kassner, e-print arXiv:cond-mat/0607823.

\bibitem{Napiorkowski} M. Napi\'orkowski and S. Dietrich, {\em Phys. Rev. E}
{\bf 47}, 1836 (1993).

\bibitem{Segovia-Lopez} J. G. Segovia-L\'opez, A. Zamora, and J. A. Santiago,
{\em J. Chem. Phys.} {\bf 135}, 064102 (2011).

\bibitem{Abate} M. Abate e F. Tovena, {\em Curve e superfici} (Springer
Italia, Milano, 2006), Observations 5.3.21 and 5.3.22.

\bibitem{Abate2} M. Abate e F. Tovena, {\em Curve e superfici} (Springer
Italia, Milano, 2006), Corollary 5.3.24.

\bibitem{Zykova2} T. Zykova-Timan, D. Ceresoli, U. Tartaglino,
and E. Tosatti, {\em J. Chem. Phys.} {\bf 123}, 164701 (2005). 

\bibitem{Milner} S. T. Milner and S. A. Safran, {\em Phys. Rev. A} {\bf 36},
4371 (1987).

\bibitem{Guenther} N. J. G\"{u}nther, D. A. Nicole, and D. J. Wallace,
{\em J. Phys. A} {\bf 13}, 1755 (1980).

\bibitem{tenWolde} P. R. ten Wolde and D. Frenkel, {\em J. Chem. Phys.}
{\bf 109}, 9901 (1998).

\bibitem{Reiss} H. Reiss and R. K. Bowles, {\em J. Chem. Phys.}
{\bf 111}, 7501 (1999).

\bibitem{Bowles} R. K. Bowles, R. McGraw, P. Schaaf, B. Senger, J.-C. Voegel,
and H. Reiss, {\em J. Chem. Phys.} {\bf 113}, 4524 (2000).

\bibitem{Maibaum} L. Maibaum, {\em Phys. Rev. Lett.} {\bf 101}, 019601
(2008).

\bibitem{Pan} A. C. Pan and D. Chandler, {\em J. Phys. Chem. B} {\bf 108},
19681 (2004).

\bibitem{Gelb} L. D. Gelb, {\em J. Chem. Phys.} {\bf 118}, 7747 (2003).

\bibitem{Mon} K. K. Mon, S. Wansleben, D. P. Landau, and K. Binder,
{\em Phys. Rev. B} {\bf 39}, 7089 (1989). 

\bibitem{Hasenbusch1} M. Hasenbusch and K. Pinn, {\em Physica A} {\bf 192},
342 (1993).

\bibitem{Rottman} C. Rottman and M. Wortis, {\em Phys. Rev. B} {\bf 29},
328 (1984).

\bibitem{Hasenbusch2} M. Hasenbusch and K. Pinn, {\em J. Phys. A} {\bf 30},
63 (1997).

\bibitem{Hoshen} J. Hoshen and R. Kopelman, {\em Phys. Rev. B} {\bf 14},
3438 (1976).

\bibitem{MuellerKrumbhaar} H. M\"{u}ller-Krumbhaar, {\em Phys. Lett. A}
{\bf 50}, 27 (1974).

\bibitem{Wulff} G. Wulff, {\em Z. Kristallogr.} {\bf 34}, 449 (1901).

\bibitem{Burton} W. K. Burton, N. Cabrera, and F. C. Frank,
{\em Philos. Trans. R. Soc. London, Ser. A} {\bf 243}, 299 (1951).

\bibitem{Li} T. Li, D. Donadio, and G. Galli, {\em J. Chem. Phys.} {\bf 131},
224519 (2009).

\bibitem{Li2} T. Li, D. Donadio, G. Russo, and G. Galli,
{\em Phys. Chem. Chem. Phys.} {\bf 13}, 19807 (2011).

\bibitem{Franke} M. Franke, A. Lederer, and H. J. Sch\"ope, {\em Soft Matter}
{\bf 7}, 11267 (2011).
\end{thebibliography}
\end{document}